\documentclass[journal=jpcbfk,manuscript=article]{achemso}
\usepackage{longtable}
\usepackage{multirow}
\usepackage{amsmath}
\usepackage[usenames,dvipsnames]{color}
\usepackage{soul}
\usepackage{threeparttable}
\usepackage{xr}
\usepackage{graphicx}
\usepackage{subfig}
\usepackage{amsmath,amssymb}
\usepackage{caption}
\usepackage{color}
\usepackage{xcolor}
\usepackage{dcolumn}
\usepackage{bm}
\usepackage{float}
\usepackage[version=4]{mhchem}
\usepackage{textcomp} \usepackage{url} \usepackage[normalem]{ulem}
\usepackage{comment}
 \usepackage{microtype}
\usepackage{etoolbox}% needed for the patch <<<
\newcommand{\ssc}[1]{\textsuperscript{#1}}

\usepackage[unicode=true, bookmarks=true, bookmarksnumbered=true,
  bookmarksopen=true, bookmarksopenlevel=2, breaklinks=false,
  pdfborder={0 0 1}, backref=false, colorlinks=true, hidelinks
]{hyperref}

\SectionNumbersOn
\author{Mitradip Das} \author{Cangtao Yin} \author{Markus Meuwly} \affiliation[University of
  Basel]{Department of Chemistry, University of Basel,
  Klingelbergstrasse 80, CH-4056 Basel, Switzerland.}
\email{m.meuwly@unibas.ch}

\title{Energy Flow Shapes the Decomposition Dynamics of a Reactive
  Atmospheric Intermediate}

\begin{document}

\date{\today}

\begin{abstract}
The unimolecular dynamics of Criegee intermediates depend strongly on
their internal energy, which can differ substantially between nascent
atmospheric species and laboratory-prepared reactants. Here, the
decomposition dynamics of {\it anti-}CH$_3$CHOO are investigated using
molecular dynamics simulations on a neural network-based potential
energy surface under thermal and mode-selective vibrational
excitation. Simulations were performed at internal energies
representative of those released during ozonolysis (chemical/thermal
preparation, TP) or following vibrational excitation (VibEx) along the
C–H stretching mode with two or three quanta, as can be done in
laboratory studies. These excitation conditions strongly affect
reaction yields and kinetics, with VibEx promoting passage across the
initial cyclization barrier. For TP, 35 kcal/mol of internal energy
are required to form molecular products, whereas for VibEx, 27
kcal/mol are sufficient. Although TP and VibEx ensembles access
largely the same fragmentation pathways with comparable relative
yields, the absolute product yields, characteristic timescales, and
kinetic behavior differ substantially. Methyldioxirane is found to be
a kinetic trap along the pathway. The resulting molecular-level
description of {\it anti-}CH$_3$CHOO decomposition provides guidance
for interpreting experiments and contributes to a broader
understanding of energy-dependent Criegee-intermediate dynamics.
\end{abstract}

\section{Introduction}
Criegee intermediates (CIs),\cite{criegee:1949} high-energy carbonyl
oxides formed during alkene ozonolysis, play an important role in
atmospheric oxidation chemistry, aerosol formation and trace-gas
budgets.\cite{marston:2008,khan:2018,chao2015direct,yin2017does,yin2018effect,yin2023nh3,yin2024revealing,cox2020evaluated}
Their subsequent fate depends sensitively on their internal energy:
nascent chemically activated CIs may undergo prompt unimolecular
reactions or be collisionally stabilized and later removed through
competing unimolecular and bimolecular
processes.\cite{Watson:2025,Robinson:2022}\\

\noindent
Following propene ozonolysis, both {\it syn}- and {\it
  anti}-CH$_3$CHOO are formed. The reaction is strongly exothermic,
releasing approximately 50 kcal
mol$^{-1}$,\cite{taatjes:2015,mcgillen:2023} but the partitioning of
this energy among the nascent products remains poorly
constrained. This is particularly relevant because
laboratory-generated CIs are typically produced from diiodo precursors
and cooled in supersonic expansions,\cite{welz:2012} leading to energy
distributions that can differ substantially from those of atmospheric
CIs.\\

\noindent
For {\it anti}-CH$_3$CHOO, the principal predicted unimolecular
pathway is cyclization to methyldioxirane.\cite{Robinson:2022}
Although bimolecular reaction with water dominates the loss of
stabilized {\it anti}-CH$_3$CHOO under humid
conditions,\cite{Lin:2016,Lin:2016.2} cyclization may become important
under drier conditions and for chemically activated CIs prior to
stabilization. Its atmospheric relevance therefore depends not only on
the barrier height, but also on the internal-energy distribution,
collisional energy transfer and competition among reaction channels.\\

\noindent
The corresponding unimolecular chemistry of {\it syn}-CH$_3$CHOO has
been investigated in considerably greater detail through spectroscopy,
kinetics measurements, imaging experiments and molecular dynamics
simulations.\cite{liu:2014,lester:2016,zhou:2019,MM.criegee:2021,MM.criegee:2023}
By contrast, {\it anti}-CH$_3$CHOO has been studied mainly through
electronic-structure calculations, statistical rate theories and
master-equation modelling,\cite{wang:2019,dyakov:2025,Robinson:2022}
and its energy-dependent cyclization dynamics remain experimentally
and dynamically unresolved.\\

\noindent
In the present study, the unimolecular dynamics of {\it
  anti}-CH$_3$CHOO are characterized on a machine-learned potential
energy surface (PES), with particular emphasis on cyclization to
methyldioxirane and the subsequent dissociation pathways. The
simulations resolve how the initial internal energy influences
reaction dynamics and product formation, providing a dynamical
description of a key unimolecular pathway of {\it anti}-CH$_3$CHOO.

\section{Results}
\begin{figure}[h!]
    \centering
    \includegraphics[width=\linewidth]{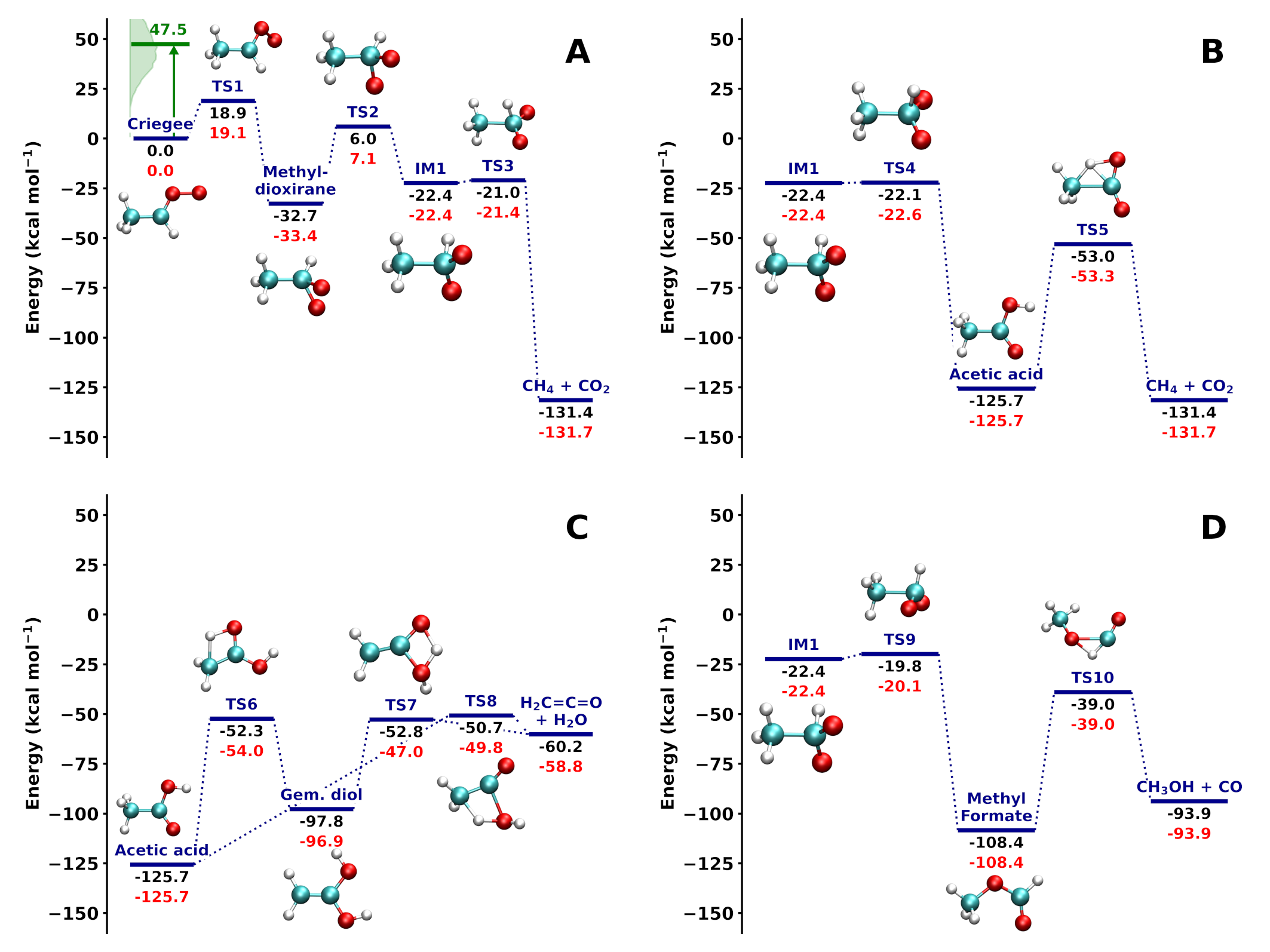}
    \caption{Major decomposition pathways of
      \textit{anti-}CH$_3$CHOO. Energies (in kcal mol$^{-1}$) from
      MP2/aug-cc-pVTZ and the ML-PES are shown in black and red
      colours, respectively. Panel A: Initial pathways for excitation
      of the Criegee intermediate to CH$_4$ + CO$_2$, via
      methyldioxirane and IM1. The green violin plot represents the
      distribution of excitation energy for one set of TP simulations
      (with $\langle E_{\rm int} \rangle = 47.5$ kcal mol$^{-1}$;
      solid green line). Panel B: Decomposition of IM1 to CH$_4$ +
      CO$_2$, via acetic acid (alternative to panel A). Panel C:
      Decomposition of acetic acid to H$_2$C=C=O + H$_2$O, either via
      geminal diol or in one step (bifurcating path in panel B). Panel
      D: Decomposition of IM1 to CH$_3$OH + CO, via methyl formate
      (connecting to IM1 in panel A).}
    \label{fig:pes}
\end{figure}

The full reaction pathway including all minima and TSs connecting
them, and the molecular products is reported in Figure \ref{fig:pes}
and described in the SI.  Along with the reaction diagram, Figure
\ref{fig:pes} demonstrates the performance of the ML-PES. For each
reactant, transition state, intermediate, and product, the reference
(MP2/aug-cc-pVTZ) and the PhysNet energies are shown in black and red,
respectively. The errors are within 2 kcal mol$^{-1}$ for all cases,
except for TS7, which is discussed further below.\\

\noindent
A yet more stringent validation involves running reactive MD
simulations and comparing NN-PES performance with electronic structure
calculations for snapshots along the trajectories; see Figure
\ref{sifig:valid-traj}. None of these structures are part of the
training set. For this analysis, 60 frames were extracted uniformly
around $\pm 0.3$ ps of each transition. The comparison of ML-PES
versus reference MP2/aug-cc-pVTZ energies shows an $R^2 > 0.999$ in
all cases except for Figure \ref{sifig:valid-traj}D, for which $R^2
\approx 0.994$. In all cases ${\rm MAE}(E) \lesssim 1$ kcal
mol$^{-1}$, demonstrating the high quality of the trained ML-PES.\\

\noindent
Next, reactive MD simulations are carried out for two different
preparations of the reactant. The first ("thermal preparation" or TP)
uses velocities from Maxwell-Boltzmann distributions for given
temperature and the second (VibEx) scales the velocity vector along
the CH-normal mode to provide internal energies typical of a
laser-based experiment. The estimated energy release of the "chemical
step" (ozonolysis of a suitable alkene) is $\sim [50 - 60]$
kcal mol$^{-1}$.\cite{cox:2020} This provides upper limits of the available
internal energy for the nascent CI. Initially, results with $\langle
E_{\rm int}^{\rm TP} \rangle = 47.5$ kcal mol$^{-1}$ are discussed which is
deliberately skewed towards the high-energy boundary of $E_{\rm int}$
to obtain statistically significant numbers of reactive
trajectories. In a second step, the conclusions are scrutinized from
additional simulations for other values of $\langle E_{\rm int}^{\rm
  TP} \rangle$. The maximum total simulation time for a single
trajectory was 1 ns throughout, which is consistent with nanosecond
time scales for prompt decomposition of chemically activated
CIs\cite{vereecken:2012} and average collision times in the
troposphere.\\

\subsection{Crossing the first barrier: \textit{anti}-CI to methyldioxirane}
First, the \textit{anti-}CI $\rightarrow$ Methyldioxirane transition
was analyzed for thermal and VibEx trajectories. For thermal
preparation (TP, $\langle E_{\rm int}\rangle \sim 47.5$ kcal mol$^{-1}$),
62.0 \% of trajectories proceed over TS1 to methyldioxirane, compared
with 69.3 \% and 18.2 \% for VibEx
with three ($E_{\rm VibEx} \sim 27.4$ kcal mol$^{-1}$) and two ($E_{\rm VibEx}
\sim 18.3$ kcal mol$^{-1}$) quanta in the CH-stretch, respectively, see Table
\ref{sitbl:final-states}. Hence, although the total internal energy
content following TP is considerably larger than for VibEx with 3
quanta, a smaller fraction of trajectories transits to methyldioxirane
on the 1 ns time scale. For VibEx with $2 \nu_{\rm CH}$ (18.3 kcal
mol$^{-1}$), still $\sim 18$ \% of the trajectories cross TS1 (18.9
kcal mol$^{-1}$ above the reactant), although excitation occurs below
the barrier. This is because the initial ensemble was generated at 300
K, and the instantaneous configurational energy together with the
vibrational energy provided allows the system to cross TS1.\\

\begin{figure}[!hbtp]
    \centering
    \includegraphics[width=\linewidth]{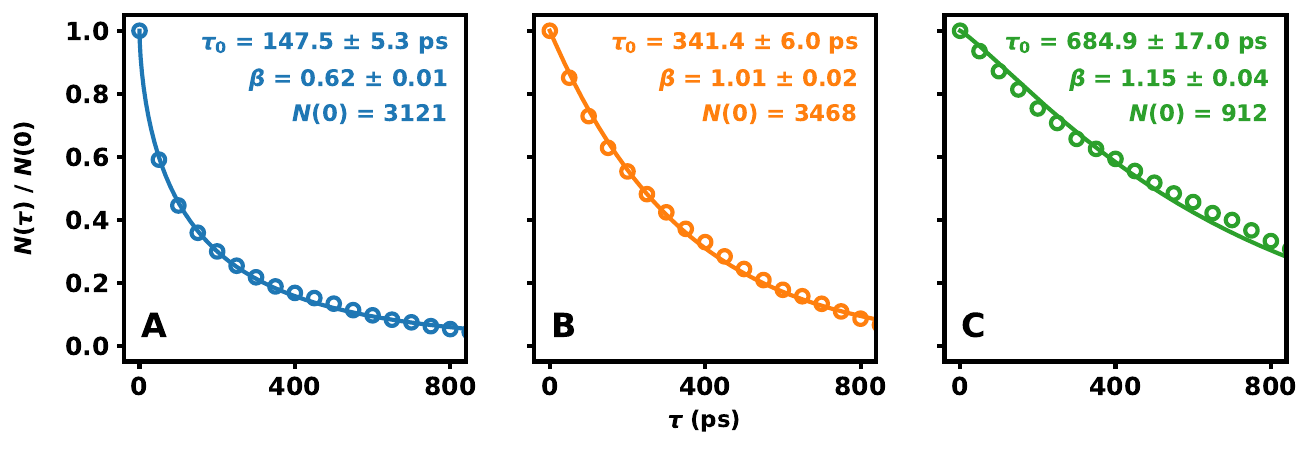}
    \caption{Survival probability $N(\tau)/N(0)$ of Criegee in all
      trajectories that proceed over TS1. Panel A: TP; panels B/C:
      VibEx preparation with 3 and 2 quanta in the CH-stretch,
      respectively. In all panels, the raw data and the fit to the
      stretched exponential function
      $\left[N(\tau)/N(0)=\exp\left(-(\tau/\tau_0)^\beta\right)\right]$
      are shown in the open circles and solid lines, respectively. The
      fitting parameters $\tau_0$ and $\beta$ are provided, along with
      $N(0)$, i.e., the total number of trajectories transiting over
      TS1. The quality of all fits are good, and errors in the fitting
      parameters were determined from
      bootstrapping.\cite{bootstrap:1979}}
    \label{fig:criegee-survival}
\end{figure}

\noindent
The survival probability $N(\tau)/N(0)$ of the reactant for all
trajectories that proceed over TS1 is reported in Figure
\ref{fig:criegee-survival}. The fit of $N(\tau)/N(0)$ versus $\tau$ to
a Kohlrausch function yields $\tau_0 \approx 147.5$ ps for TP. On the
other hand, $\tau_0 \approx 341.4$ ps and $684.9$ ps for VibEx with
$3\nu_{\rm CH}$ and $2\nu_{\rm CH}$, respectively. Additionally, TP
features $\beta \approx 0.62$ from the stretched exponential fit,
pointing to a distribution of timescales. In contrast, VibEx gives
$\beta \approx 1$ for $3\nu_{\rm CH}$ excitation, consistent with
single-exponential decay, and $\beta \approx 1.15$ for $2\nu_{\rm CH}$
excitation, indicative of a slightly compressed-exponential
dynamics. Hence, the efficiency to cross TS1 is largest for $3\nu_{\rm
  CH}$ whereas the kinetics for TP is at least 2.5 times faster than
for VibEx.\\

\subsection{Methyldioxirane to Products: Yields and Kinetics}
After reaching methyldioxirane, a fraction of the trajectories lead to
fragmentation products, see Table
\ref{sitbl:final-states}. Unimolecular dissociation to molecular
products occurs for 26.6 \%, 8.6 \%, and 0.5 \% of all (5000)
trajectories. Hence, while $\sim 42$ \% of all TP-trajectories
reaching methyldioxirane proceed to products, only $\sim 12$ \% and
$\sim 3$ \%, respectively, do so for VibEx trajectories on the 1 ns
time scale. This indicates that post-methyldioxirane the energy in the
modes for VibEx trajectories has been redistributed, as further
transition to IM1 takes place through a different reaction coordinate
than \textit{anti}-CI $\to$ methyldioxirane. The comparatively large
fraction of TP-trajectories that dissociates is probably related to
the considerably larger energy content of the reactant state. Given
that for VibEx only 15 \% or fewer of the trajectories move to
molecular products indicates that methyldioxirane can act as a kinetic
trap. This effect is less pronounced for TP.\\

\begin{table}[h!]
    \centering
    \begin{tabular}{cc|c|cc}
        \hline
         Product & Precursor & TP  & VibEx (3$\nu_{\rm CH}$) & VibEx (2$\nu_{\rm CH}$) \\
         \hline
         \multirow{3}{*}{CH$_4$ + CO$_2$}
         & Methyldioxirane/IM1 & $37.1 \pm 1.3$ \% & $41.0 \pm 2.4$ \% & $31.2 \pm 8.8$ \%\\
         & Acetic acid &       $11.4 \pm 0.9$ \% & $11.6 \pm 1.5$ \% & $27.1 \pm 8.8$ \%\\
         & \textbf{Total} &    $49.1 \pm 1.3$ \% & $52.7 \pm 2.3$ \% & $57.5 \pm 9.8$ \%\\
         \hline
         \multirow{3}{*}{H$_2$C=C=O + H$_2$O}
         & Acetic acid &    $32.8 \pm 1.3$ \% & $36.1 \pm 2.3$ \% & $26.6 \pm 8.8$ \% \\
         & Geminal diol &   $2.0 \pm 0.4$ \%  & $3.0 \pm 0.8$ \%  & Not observed\\
         & \textbf{Total} & $36.0 \pm 1.4$ \% & $39.9 \pm 2.4$ \% & $26.6 \pm 8.8$ \% \\
         \hline
         \multirow{2}{*}{CH$_3$OH + CO}
         & Methyl formate & $10.0 \pm 0.8$ \% & $7.0 \pm 1.2$ \% & $15.5 \pm 7.2$ \%\\
         & \textbf{Total} & $10.0 \pm 0.8$ \% & $7.0 \pm 1.2$ \% & $15.5 \pm 7.2$ \%\\
         \hline
    \end{tabular}
    \caption{Major fragmentation products, along with their immediate
      precursors, from MD simulations for TP and VibEx
      preparation. The errors are computed using
      bootstrapping.\cite{bootstrap:1979} Some minor products and/or pathways are not
      reported.}
    \label{tbl:frag-products}
\end{table}

\noindent
The fragmentation product distribution for all three products (CH$_4$
+ CO$_2$, H$_2$C=C=O + H$_2$O, and CH$_3$OH + CO) involving the three
preparations is summarized in Table \ref{tbl:frag-products}. About 50
\% of the trajectories end up in CH$_4$ + CO$_2$. This state is
primarily reached directly via decomposition of methyldioxirane/IM1
through TS3. The second most populated product is H$_2$C=C=O + H$_2$O,
which forms through decomposition of acetic acid via TS8 in all cases,
whereas a smaller percentage proceeds via the geminal diol. Note that,
since the ML-PES overestimates the energy of TS7, the yields of
H$_2$C=C=O + H$_2$O through the geminal diol should be considered a
lower limit. This point is discussed further below. Finally, the
lowest percentage of trajectories generates CH$_3$OH + CO via methyl
formate. Overall, it is found that all preparations yield the same
{\it ranking} whereas the percentages of flux depends on the initial
preparation.\\

\begin{figure}[!hbtp]
    \centering
    \includegraphics[width=\linewidth]{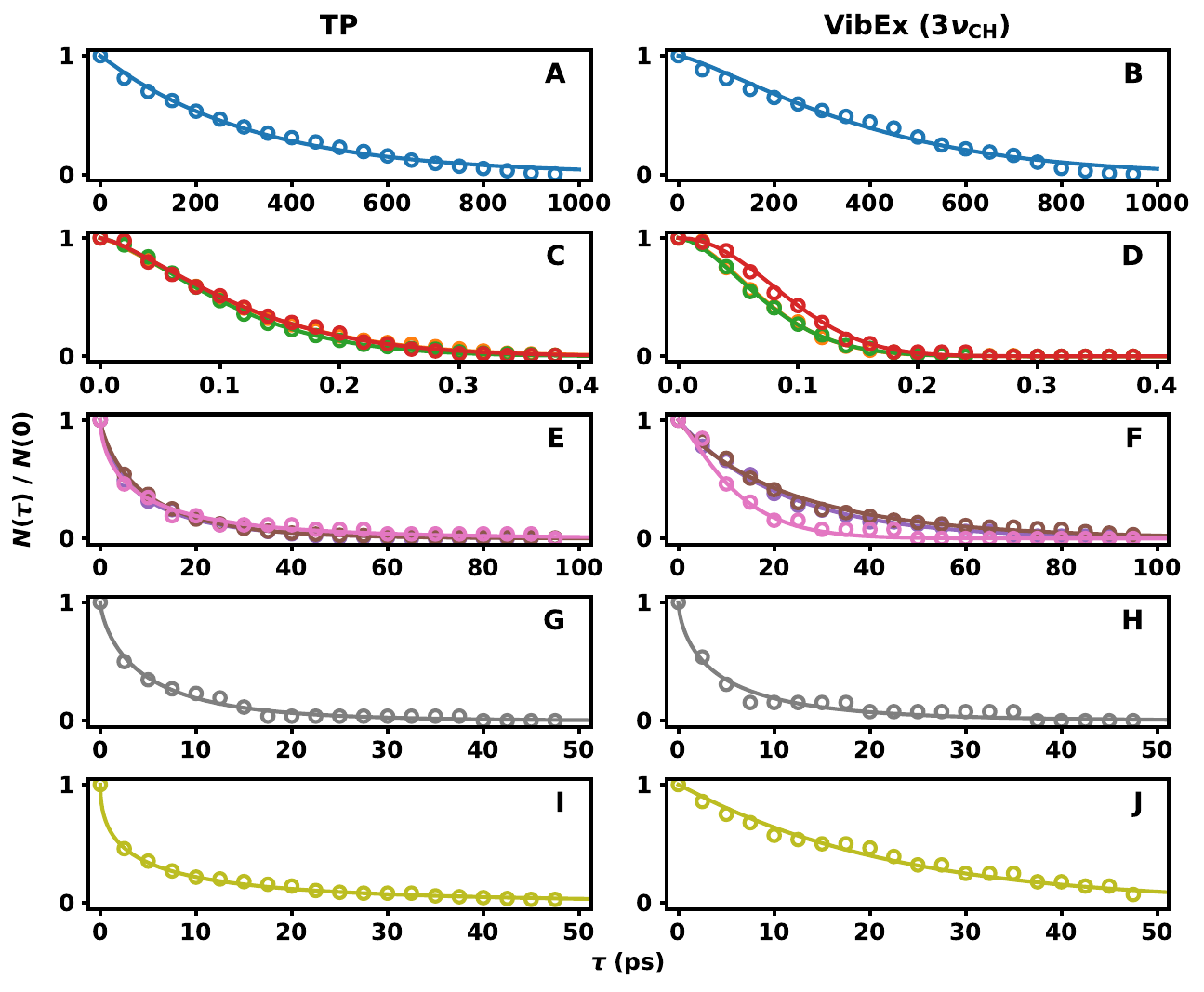}
    \caption{Survival probabilities for various intermediates
      corresponding to individual reaction channels. Panels A, B:
      Methyldioxirane $\rightarrow$ IM1 (blue). Panels C, D: IM1
      $\rightarrow$ [CH$_4$ + CO$_2$ (orange), acetic acid (green),
        methyl formate (red)]. Panels E, F: Acetic acid $\rightarrow$
      [CH$_4$ + CO$_2$ (violet), H$_2$C=C=O + H$_2$O (brown), Geminal
        diol (magenta)]. Panels G, H: Geminal diol $\rightarrow$
      H$_2$C=C=O + H$_2$O (gray). Panels I, J: Methyl formate
      $\rightarrow$ CH$_3$OH + CO (lime). Panels A, C, E, G, and I
      correspond to TP, while panels B, D, F, H, and J are for VibEx
      with 3 quanta of energy. In all panels, the open circles and the
      solid lines correspond to the raw data and the fit to a
      stretched exponential function
      $\left[N(\tau)/N(0)=\exp\left(-(\tau/\tau_0)^\beta\right)\right]$. The
      fitting parameters are summarized in Table \ref{tbl:frag-kin}.}
    \label{fig:frag-kin}
\end{figure}

\noindent
Survival probabilities $N(\tau)/N(0)$ for every intermediate along
each of the fragmentation channels for TP and VibEx ($3 \nu_{\rm CH}$)
are reported in Figure \ref{fig:frag-kin}. The survival for the VibEx
pathway with 2 quanta of energy was not computed since only 0.5 \% of
(i.e. 26) trajectories lead to fragmentation. Again, $N(\tau)/N(0)$
versus $\tau$ were fit to a Kohlrausch function. Clearly, in both
cases, the kinetic bottleneck involves the transition from
methyldioxirane to IM1. This process is slower than the first
transition from {\it anti-}CI $\to$ methyldioxirane (Figure
\ref{fig:criegee-survival}), despite its lower energy barrier compared
to the methyldioxirane $\to$ IM1 transition (Figure \ref{fig:pes}). In
addition, the long characteristic timescales for methyldioxirane $\to$
IM1 transition compared to \textit{anti}-CI $\to$ methyldioxirane
transition support the possibility of a kinetic trap at
methyldioxirane for the decomposition pathways. This is further
supported by the extended mean residence times of $\sim 294$ ps for TP
and $\sim 369$ ps for VibEx ($3\nu_{\rm CH}$), both substantially
exceeding the {\it anti}-CI decomposition timescale and approaching
the characteristic timescale of methyldioxirane dynamics, see Figure
\ref{sifig:res-time-dioxi}.\\

\begin{table}[!hbtp]
    \centering
    \begin{tabular}{ll|cc|cc}
      \hline
         \multirow{2}{*}{Initial} & \multirow{2}{*}{Final} & \multicolumn{2}{c|}{TP} & \multicolumn{2}{c}{VibEx ($3\nu_{\rm CH}$)}
         \\
         
         & & $\tau_0$ (ps) & $\beta$ & $\tau_0$ (ps) & $\beta$\\
         \hline
         Methyldioxirane & IM1 & $316.12 \pm 10.0$ & $1.01 \pm 0.0$ & $422.76 \pm 18.5$ & $1.27 \pm 0.1$ \\
         \hline
         \multirow{3}{*}{IM1} 
         & CH$_4$ + CO$_2$ & $0.13 \pm 0.0$ & $1.32 \pm 0.1$ & $0.09 \pm 0.0$ & $1.72 \pm 0.1$ \\
         & Acetic acid     & $0.12 \pm 0.0$ & $1.43 \pm 0.1$ & $0.08 \pm 0.0$ & $1.68 \pm 0.1$ \\
         & Methyl formate  & $0.13 \pm 0.0$ & $1.38 \pm 0.1$ & $0.11 \pm 0.0$ & $2.07 \pm 0.4$ \\
         \hline
         \multirow{3}{*}{Acetic acid} 
         & CH$_4$ + CO$_2$       & $8.46 \pm 1.1$ & $0.72 \pm 0.1$ & $21.88 \pm 3.2$ & $0.99 \pm 0.1$\\
         & H$_2$C=C=O + H$_2$O & $9.81 \pm 0.6$ & $0.79 \pm 0.0$ & $23.86 \pm 2.3$ & $0.91 \pm 0.1$\\
         & Geminal diol          & $8.65 \pm 3.3$ & $0.60 \pm 0.1$ & $12.64 \pm 3.1$ & $1.21 \pm 0.5$\\
         \hline
         Geminal diol & H$_2$C=C=O + H$_2$O & $4.90 \pm 1.6$ & $0.71 \pm 0.2$ & $4.56 \pm 2.3$ & $0.65 \pm 0.4$\\
         \hline
         Methyl formate & CH$_3$OH + CO & $4.30 \pm 1.1$ & $0.49 \pm 0.0$ & $21.78 \pm 4.4$ & $1.03 \pm 0.2$\\
         \hline
    \end{tabular}
    \caption{Stretched exponential fit to the survival probabilities,
      i.e. $\left[N(\tau)/N(0)=\exp\left(-(\tau/\tau_0)^\beta\right)\right]$,
      for different species in each reaction channel, for both TP and
      VibEx trajectories. $\tau_0$ and $\beta$ correspond to the
      characteristic timescale and stretching exponent,
      respectively. The errors are computed with
      bootstrapping.\cite{bootstrap:1979}}
    \label{tbl:frag-kin}
\end{table}

\noindent
The next step, involving IM1 branching to CH$_4$ + CO$_2$, acetic
acid, or methyl formate, involves a low barrier and therefore takes
place on sub-picosecond timescales. This step features compressed
exponential behavior ($\beta > 1$) for both TP and VibEx. The
elementary reactions in Table \ref{tbl:frag-kin} show a maximum
$\tau_0 \sim 10$ ps for TP compared with $\sim 25$ ps for VibEx,
except for the methyldioxirane $\to$ IM1 step. In reactions from IM1
to fragmentation products, all elementary reaction steps for TP
feature $\beta<1$, indicating a distribution of relaxation timescales
in the system. On the other hand, for VibEx, all elementary steps in
acetic acid $\rightarrow$ CH$_4$+CO$_2$, acetic acid $\rightarrow$
H$_2$C=C=O + H$_2$O and methyl formate $\rightarrow$ CH$_3$OH + CO
yield $\beta \approx 1$. In contrast, the acetic acid $\rightarrow$
geminal diol $\rightarrow$ H$_2$C=C=O + H$_2$O pathway features $\beta
\approx 1.2$ and $\beta \approx 0.6$, respectively.\\

\subsection{\textit{anti}-CI to Products: Overall reaction kinetics}
\begin{figure}[!hbtp]
    \centering
    \includegraphics[width=\linewidth]{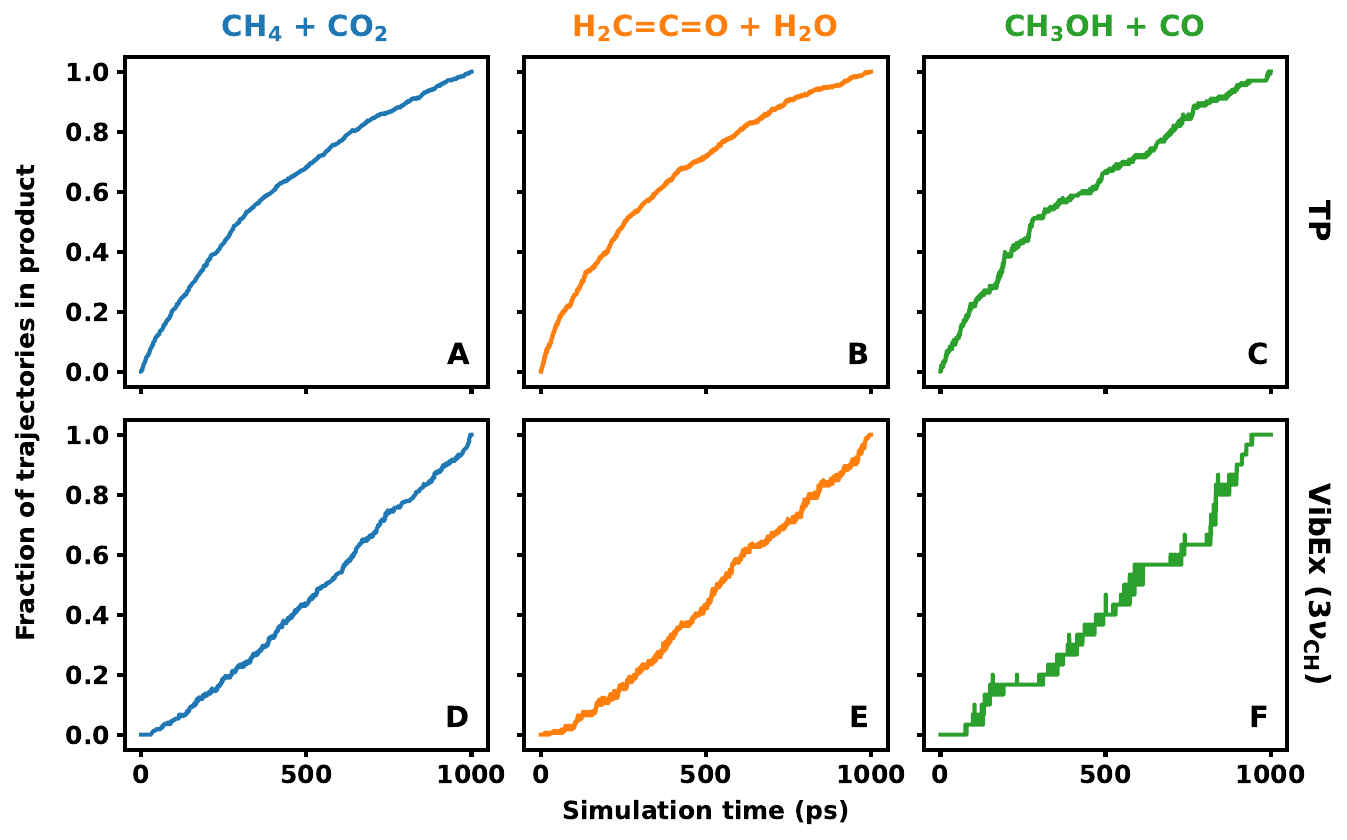}
    \caption{Fraction of the products formed as a function of
      simulation time for TP (panels A-C at 2000 K) and VibEx (panels
      D-F, 3 quanta). Products formed include CH$_4$ + CO$_2$ (panels
      A/D), H$_2$C=C=O + H$_2$O (panels B/E), and CH$_3$OH + CO
      (panels C/F). The dashed black lines indicate the time
      $\tau_{1/2}$ at which 50\% of the trajectories have formed
      products. In addition, the initiation time $\tau_\text{init}$ is
      also reported, denoting the time before the first product
      formation.}
    \label{fig:product}
\end{figure}

Finally, the overall kinetics for product formation starting from the
energized reactant is discussed. Figure \ref{fig:product} shows the
fraction of trajectories forming specific products as a function of
time, along with values for $\tau_{1/2}$ and $\tau_\text{init}$. The
value of $\tau_{1/2}$ indicates the timescales required for 50 \% of
the trajectories to form products, whereas $\tau_\text{init}$ is the
initiation time, i.e., the time required for the first product to
appear.\\

\noindent
The kinetic traces for product formation, see Figure \ref{fig:product}
top (thermal) and bottom (VibEx) panels, differ markedly. For TP, the
curves show a non-linear behaviour, whereas for VibEx an almost linear
increase is observed. For both preparations, the behaviour deviates
clearly from first-order kinetics, see Figure \ref{sifig:product-log}
which reports $\log{(N(\tau)/N(0))}$ vs. $t$. As Figure
\ref{fig:product}A/B/C shows, $\tau_{\rm init} < 1.5$ ps for TP,
whereas for VibEx (Figure \ref{fig:product}D/E/F), $\tau_\text{init}
\sim 10-80$ ps. This is related to intramolecular vibrational
redistribution (IVR) between the excited CH-stretch and the reaction
coordinate leading across TS1 that needs to take place for VibEx
preparation.  Similarly, the values for $\tau_{1/2}$ for the two
preparations differ by a factor of $\sim 2$: for the TP they are
typically 250 to 300 ps and increase to $\sim$ 550 ps for VibEx. Since
there are only 28 trajectories in Figure \ref{fig:product}F leading to
CH$_3$OH+CO, the concentration curve is rather jagged.\\

\noindent
Methyldioxirane decomposition (Figure \ref{sifig:product-dioxi}) is
characterized by $\tau_0 \sim 300$ ps (TP) and $\sim 450$ ps (VibEx),
both comparable to the corresponding $\tau_{1/2}$ for product
formation. On the other hand, with \textit{anti}-CI, the corresponding
$\tau_0$ for decomposition are $\sim 40 $ ps and $\sim 130$ ps, for TP
and VibEx, respectively; see Figure \ref{sifig:product-criegee}. These
timescales are 5 to 8-fold shorter than the product-formation
$\tau_{1/2}$. Further, a fast rise followed by a slow decay can be
observed in the time-dependent fraction of methyldioxirane (Figure
\ref{sifig:dioxi-fraction}) with a local maximum at $\sim 100$
ps. This is consistent with accumulation of species B (here
Methyldioxirane) in a consecutive reaction scheme $A
\overset{k_1}{\to} B \overset{k_2}{\to} C$ with $k_1 > k_2$, for which
there is a fast rise (A $\to$ B), followed by a slow decay (B $\to$
C), and a local maximum for [B] in between.\cite{marin:2010}\\

\section{Discussion and Conclusions}
In this study, the possible reaction pathways for the CH$_3$CHOO {\it
  anti-}CI intermediate through unimolecular reaction dynamics were
characterized with both, TP ($\langle E_{\rm int}\rangle \sim$ 47.5
kcal mol$^{-1}$) and VibEx with $3\nu_{\rm CH}$ ($E_{\rm VibEx}\sim
27.4$ kcal mol$^{-1}$) and $2\nu_{\rm CH}$ ($E_{\rm VibEx} \sim 18.3$
kcal mol$^{-1}$). For TP, $\sim 27\%$ of all trajectories undergo
fragmentation, with $\tau_{1/2} \sim 250-300$ ps. In contrast, VibEx
excitation ($3\nu_{\rm CH}$) yields fragmentation in only $\sim 9\%$
of all trajectories, with $\tau_{1/2} \sim 530-560$ ps. However, the
major relative product yields in all cases are 
   $ \sim 50\%$, $\sim 40\%$ and $\sim 10\%$ for
CH$_4$ + CO$_2$, H$_2$C=C=O + H$_2$O, and CH$_3$OH + CO, irrespective
of the type of preparation. Across the elementary steps, $\tau_0$
varies by up to a factor of 2.5 between TP and VibEx. The kinetics are
predominantly stretched exponential $(\beta < 1)$ for TP and
compressed exponential $(\beta > 1)$ for VibEx. Thus, although the
relative importance of the fragmentation pathways remains comparable,
the total product yields, characteristic time scales, and even the
character of the kinetics inferred from fits to a Kohlrausch function
depend substantially on the initial preparation.\\

\noindent
In an attempt to further improve the ML-PES for the TS7 energy, an
additional NN-PES was trained. For this new, retrained model, 1000
structures were generated near TS7 and along the IRC connecting
geminal diol and H$_2$C=C=O + H$_2$O using NMS and RDS, respectively,
and were added to the training set. For the retrained model the energy
of TS7 decreased to $-52.9$ kcal mol$^{-1}$, close to the reference
value of $-52.8$ kcal mol$^{-1}$ and an improvement over $-47.0$ kcal
mol$^{-1}$ from the ML-PES. Next, the final states (Table
\ref{sitbl:retrain-final-states}), the fragmentation product yields
(Table \ref{sitbl:retrain-frag-products}), and the kinetics of
elementary steps (Table \ref{sitbl:retrain-frag-kin}) were analyzed
from 1000 trajectories with $\langle E_{\rm int} = 53.5 \rangle$ kcal
mol$^{-1}$. Overall, results between the original and retrained model
agree well, including processes involving TS7. For example, in both
models, $\sim 25$ \%, $\sim 35$ \% and $\sim 39$ \% of all
trajectories end up in {\it anti-}CI, methyldioxirane and
fragmentation products at the end of 1 ns; see Table
\ref{sitbl:retrain-final-states}. In addition, among the trajectories
forming fragmentation products, $\sim$ 50 \%, $\sim$ 35 \% and $\sim$
8 \% form CH$_4$ + CO$_2$, H$_2$C=C=O + H$_2$O, and CH$_3$OH + CO,
respectively, for both models; see Table
\ref{sitbl:retrain-frag-products}. However, the fraction of
trajectories in acetic acid $\to$ CH$_4$ + CO$_2$ and \ce{H2C=C=O} +
H$_2$O reduces by $\sim4$ \%, and the share of geminal diol $\to$
\ce{H2C=C=O} + H$_2$O increases by $\sim 4$ \%. In conclusion, the
ML-PES used for the majority of the simulations is still considered
sufficiently reliable for dynamics studies.\\

\noindent
Since the precise energy content of CI after the chemical step
(ozonolysis) is unknown, it is relevant to also consider the yields
and kinetics for a range of internal energy contents. For this, MD
simulations were run with $\langle E_{\rm int} \rangle = 29.8$,
$\langle E_{\rm int} \rangle = 41.8$ and $\langle E_{\rm int} \rangle
= 53.5$ kcal mol$^{-1}$, respectively, see Figures \ref{sifig:int-e}
and \ref{sifig:eint}, compared with an estimated total energy release
upon ozonolysis of up to 60 kcal mol$^{-1}$.\cite{cox:2020} The
probability to form fragmentation products for given $\langle E_{\rm
  int} \rangle$ considered here is reported in Figure
\ref{sifig:frag-prob}. With $\langle E_{\rm int} \rangle = 41.8$ kcal
mol$^{-1}$, 46 \% of the trajectories traverse TS1 to form
methyldioxirane, 13 \% of which move on towards products. These
fractions change to 12 \%, and 1.2 \% at $\langle E_{\rm int} \rangle
= 29.8$ kcal mol$^{-1}$, respectively (Table
\ref{sitbl:final-states}). On the other hand, within error bars the
relative percentages to form CH$_4$ + CO$_2$, H$_2$C=C=O + H$_2$O, and
CH$_3$OH + CO products are largely insensitive to $\langle E_{\rm int}
\rangle$, see Table \ref{sitbl:frag-products}. The characteristic
timescales for the decay of \textit{anti}-CI and the elementary steps
for the formation of products are inversely correlated with the
internal energy content, as shown in Figure \ref{sifig:frag-prob} and
Table \ref{sitbl:final-states}.\\

\noindent
The analysis so far demonstrates that for TP at least $E_{\rm int} =
35$ kcal mol$^{-1}$ is required to form products, see Figure
\ref{sifig:eint}. This is a plausible amount for $E_{\rm int}$ after
the "chemical step". Hence, to probe scenarios relevant to the
chemistry in the troposphere, the 250 trajectories with lowest $E_{\rm
  int}$ from the full body of 12000 TP-trajectories that form
molecular products was analyzed. The fragmentation products for this
``mixed dataset'' are reported in Table \ref{sitbl:frag-products} and
the kinetic analysis is in Table \ref{sitbl:frag-kin}. The yields and
Kohlrausch parameters are consistent with the results for $E_{\rm int}
= 47.5$ kcal mol$^{-1}$, see Table \ref{tbl:frag-products} ([46,32,9]
\% vs. [49,36,10] \%), and appear to not depend significantly on the
way how the TP-initial conditions were generated. Subsampling 1000
times from 250 samples of the $\sim 1500$ reactive trajectories with
$\langle E_{\rm int} \rangle = 47.5$ kcal mol$^{-1}$ shows that 250
reactive trajectories are representative of a considerably larger
ensemble (Table \ref{sitbl:frag-products}), although the error bars
from fewer trajectories clearly increase.\\

\noindent
Among the reaction pathways for decomposition of {\it anti-}CI, some
pathways, such as formation of CH$_4$ + CO$_2$ via IM1 or acetic acid,
and the formation of H$_2$C=C=O + H$_2$O via acetic acid, were
previously reported.\cite{dyakov:2025} On the other hand, the pathway
to H$_2$C=C=O + H$_2$O via geminal diol and to CH$_3$OH + CO via
methyl formate appear to be newly identified in the current
study. Some previously reported products including CH$_3$ + COOH and
CH$_3$CO + OH were not identified in the present work since these are
high-energy species.\cite{dyakov:2025} Importantly, methyldioxirane
was identified as a kinetic trap in the present case, which was not
the case for the decomposition of H$_2$COO.\cite{MM.h2coo:2025}\\

\noindent
Finally, the influence of multi-reference effects was briefly
considered. For all local minima and transition states, CASPT2
energies using the aug-cc-pVTZ basis set were determined. The first
barrier (TS1) differs only by 0.5 kcal/mol whereas the ensemble of
subsequent transition states are uniformly higher in energy by 10
kcal/mol, see Table \ref{sitab:multi-ref} and Figure
\ref{sifig:mr}. The corresponding intermediates (methyldioxirane, IM1,
acetic acid, and methyl formate) are also less stable by ca. 10
kcal/mol except for the geminal diol, for which the difference is only
3 kcal/mol. Since the majority of states downstream from TS1 are
shifted by a constant amount, the relative yields of the fragmentation
products will not be affected. However, because methyldioxirane at the
MP2-level is $\sim 50$ kcal/mol below TS1 compared with 44 kcal/mol at
CASPT2, the number of trajectories forming molecular products on the 1
ns time scale may decrease.\\

\noindent
The present work shows that the unimolecular decomposition of
\textit{anti}-CH$_3$CHOO is strongly preparation-dependent. The
reaction dynamics under conditions (TP) typical for the troposphere
differs substantially from that initiated by laboratory-based
vibrational excitation (VibEx). TP-conditions that are probably most
representative are nonequilibrium (multi-temperature) distributions
with internal energies in a particular energy window (here $35 <
E_{\rm int} < 53$ kcal mol$^{-1}$). Neither truly single-temperature
Maxwell-Boltzmann nor microcanonical conditions (constant $E_{\rm
  int}$) are expected to provide realistic initial
preparations. Although TP and VibEx ensembles access largely the same
fragmentation pathways with comparable relative importance, their
absolute product yields, characteristic timescales, and kinetic
behavior differ substantially. Thus, similarity in reaction pathways
does not necessarily imply similarity in reaction kinetics or product
formation.\\

\noindent
Viewed more broadly, the present findings highlight the need for
caution when transferring reaction parameters derived from laboratory
experiments directly into reaction-network models of the chemical
evolution of terrestrial or astrophysical atmospheres. As
high-accuracy representations of reactive molecular potential energy
surfaces become increasingly available, systematic exploration of the
underlying reaction dynamics is becoming feasible on a routine
basis. Such simulations, validated against laboratory measurements
whenever possible,\cite{MM.nno:2023} can help bridge the gap between
molecular-level dynamics and atmospheric-scale reaction models,
particularly under conditions where direct experimental
characterization is difficult or impossible.\\

\section{Methods}

\subsection{Training the NN-PES}
All electronic structure calculations were carried out at the
second-order M{\o}ller-Plesset perturbation theory (MP2) with a
correlation-consistent augmented polarised triple-zeta (aug-cc-pVTZ)
basis set.\cite{moller1934note,gordon88mp2,dunning89BtoNe} The
Gaussian 16 program was employed to obtain optimized structures for
the intermediates and the transition states, as well as the structures
along the IRC pathways.\cite{g16} Energies, forces, and dipole moments
for generating the ML dataset were calculated using Molpro
2024.1.\cite{molpro:2020} Energies for all structures were referenced
to the optimized structure for \textit{anti-}Criegee, which was the
zero of energy.\\

\noindent
A NN-PES was employed to obtain the full description of the energy
landscape connecting \textit{anti-}CH$_3$CHOO to various reaction
intermediates, transition states separating them, and their
decomposition products. For intermediates, the local structure of the
energy landscape was captured using molecular dynamics (MD)
simulations with the extended tight-binding model (XTB) simulated in
the Atomic Simulation Environment (ASE), as well as using normal mode
sampling (NMS) computed using in-house
code.\cite{ase:2017,xtb.grimme:2021} For each intermediate (local
minimum), the training set contained $\sim 1000$ structures from MD
simulations and $\sim 1000$ from NMS.  For each TS, $\sim 1000$ NMS
structures were generated to characterize the local PES, whereas $\sim
1000$ structures were generated using Random Distribution Sampling
(RDS) along each forward and reverse IRC. For this, the Cartesian
coordinates of each of the atoms were randomly displaced by $\pm 0.05$
\AA\/, $\pm 0.10$ \AA\/, and $\pm 0.40$ \AA\/, respectively, from the
points along the IRC, with 20 \%, 40 \% and 40 \% of the structures
assigned to each displacement range, respectively. This yielded a
total of $\sim 4.8 \times 10^4$ structures, used for initial training.\\

\noindent
This first data set was extended through 13 rounds of adaptive
sampling, leading to an additional $\sim 2.6 \times 10^4$ structures
in the training set. The primary purpose here was to eliminate all
spurious minima to obtain a stable, global NN-PES suitable for MD
simulations. The final data set contained $\sim 7.4 \times 10^4$
structures for which energies, forces, and dipole moments were
available for training.\\

\noindent
For training the NN-PES, $\sim 4.3 \times 10^3$ high-energy structures
with a total energy 8 eV ($\sim$ 184.5 kcal mol$^{-1}$) or more above
the minimum energy of the reactant were removed since this energy
range was not relevant for the dynamics studies. The remaining $\sim 7
\times 10^4$ structures were split according to 80/10/10 into
training, validation, and test data. PhysNet, implemented with
TensorFlow 2.21, was employed to train the
NN-PES.\cite{MM.physnet:2019,physnet-tf2} The input layer of PhysNet
involves atomic numbers and atom positions, whereas the output layer
predicts the energy, atomic forces, the total charge, and the total
dipole moment. The loss function was
\begin{equation}
    \mathcal{L} = w_{E} \mathcal{L}_E + \frac{w_{F}}{3 N} \mathcal{L}_F 
    +w_Q \mathcal{L}_Q + w_P \mathcal{L}_P + \mathcal{L}_{nh}
    \label{eqn:loss}
\end{equation}
whereby $\mathcal{L}_X$ denotes the L1 loss and $w_E \approx 1$,
$w_F\approx 53$, $w_Q \approx 14$ and $w_P\approx 27$, are the
relative weights of each contribution to the total loss. The
non-hierarchical penalty $\mathcal{L}_{nh} (\approx 0.01)$ prevents
overfitting of the model. Four individual models, with unique random
seeds and dispersion switched off, were trained for $10^6$ steps each
on NVIDIA RTX 3080 and RTX 3090 GPUs. In all cases, the AMSGrad
optimizer was employed to optimize the loss function with a learning
rate of $10^{-4}$. Additional technical details on the architecture of
PhysNet and the hyperparameters are provided in the supporting
information and in Table \ref{sitbl:hyper}.  \\

\noindent
To assess model quality, the mean absolute (MAE) and root mean squared
(RMSE) errors for energy and forces were computed on the test set
($\sim 7000$ structures). All trained models were ranked by ${\rm
  MAE}(E)$. The best model (lowest MAE of energy) was further
evaluated using diffusion Monte-Carlo (DMC) simulations to check for
``holes'', starting simulations from all
intermediates.\cite{dmc:1996,dmc.bowman:2021,MM.atmos:2020} Following
this, the best-performing model was chosen to carry out reactive MD
simulations. It is noted that the forces were used for training
purposes only. For the MD simulations (see below), forces were
determined from the energy through reverse-mode automatic
differentiation and not from evaluating the NN-PES.\\

\noindent
The statistical benchmarks for all four models are provided in Table
\ref{sitbl:test-data}. For all models, the overall ${\rm MAE}(E) \sim
1$ kcal mol$^{-1}$ and $R^2 > 0.99$ on the test set with respect to
the MP2/aug-cc-pVTZ reference data. Within the energy range $E \in
[-135,35]$ kcal mol$^{-1}$, which is sampled during the reactive MD
simulations (shaded area in Figure \ref{sifig:valid-test}A), the
performance improves to ${\rm MAE}(E) < 0.5$ kcal mol$^{-1}$. The drop
in performance when using the entire data set arises from high-energy
structures; see Figure \ref{sifig:valid-test}A.\\

\noindent
The PhysNet-4 model (``ML-PES'' for the remainder of the work, bold in
Table \ref{sitbl:test-data}) was taken forward for simulations based
on the lowest ${\rm MAE}(E)$ and ${\rm RMSE}(E)$ over the full test
set among the models trained. The quality of the ML-PES on the test
dataset is shown in Figure \ref{sifig:valid-test}. Clearly, data
points with large energy mismatch cluster in the high-energy region of
the PES, which is not relevant for the present MD simulations (shaded
area). Despite covering an energy range of $\sim 300$ kcal mol$^{-1}$,
for 90 \% of the test data a performance of $|\Delta E| < 1$ kcal
mol$^{-1}$ between reference data and the NN-PES was achieved.\\

\subsection{Molecular Dynamics Simulations}
All reactive MD simulations were carried out using the ASE software in
an $NVE$ ensemble, using a time step $\Delta t = 0.1$ fs and a total
simulation time for each trajectory of 1 ns.\cite{ase:2017} For
analysis, frames were written every 10 fs. For each type of
simulation (see below), 5000 independent simulations were carried
out.\\

\noindent
To follow thermal decomposition, the simulations were initialized from
the optimized reactant structure with velocities sampled from a
Maxwell-Boltzmann distribution at 2000 K, providing a total internal
energy of $\sim 47.5$ kcal mol$^{-1}$ on average. On the other hand,
to investigate vibrationally excited (VibEx) decomposition, first a 1
ns simulation starting from the Criegee intermediate in the constant
$NVT$ ensemble at 300 K was carried out, employing Langevin equations
of motion. Next, 5000 structures from the ensemble were selected
randomly, and the CH bond was excited by scaling the atom velocities
along the CH-normal mode to contain the equivalent of either 2- or
3-CH-quanta, respectively, corresponding to $E_{\rm VibEx} \sim $ 18.3
kcal mol$^{-1}$ and $\sim$ 27.4 kcal mol$^{-1}$ of energy, following
earlier work.\cite{MM.criegee:2023,MM.h2coo:2026}\\

\noindent
{\it Structure Assignment:} In-house Python codes, employing NumPy,
SciPy, and NetworkX libraries, were used to analyze all
trajectories.\cite{numpy,scipy,networkx} For each analyzed structure,
a graph was constructed using the atoms as nodes and bonds as edges. A
bond between atoms A and B was assigned if $r_{\rm AB}<1.5 \times
(R_{\rm A} + R_{\rm B})$, where $r_{\rm AB}$ denotes the distance
between the atoms, and $R_{\rm A}$ and $R_ {\rm B}$ are their
respective covalent radii. Here, $R_\text{C} = 0.76$ \AA, $R_\text{O}
= 0.66$ \AA\ and $R_\text{H} = 0.31$ \AA\ was
used. \cite{pauling_book:1960, cov.radii:2008} The scaling factor
$1.5$ was introduced to account for bond vibrations. The number of
connected components in the graph was used to determine fragmentation,
where the products are identified based on their stoichiometric
formula. In case of no fragmentation in a frame, a graph isomorphism
test, allowing permutation between nodes corresponding to the atoms of
the same element, between the frame and the list of intermediates and
TS, was employed to identify the molecular species for a given
MD-frame. Thus, this provides a structurally resolved view of the
trajectory, which allows one to classify it into fragmentation
pathways.\\

\noindent
\textit{The survival probability} $N(\tau)/N(0)$ was computed to
characterize the kinetics between two given states A and B. Here,
$N(0)$ refers to the total number of trajectories involved in the
process considered, and $N(\tau)$ is the number of trajectories that
remain in state A after time $\tau$. The resulting curve
$N(\tau)/N(0)$ versus $\tau$ was fitted to a Kohlrausch function
(``stretched exponential'')
$\left[N(\tau)/N(0)=\exp\left(-(\tau/\tau_0)^\beta\right)\right]$,
where $\tau_0$ and $\beta$ represent a characteristic timescale and
stretching exponent, respectively. Error estimation for the fitting
parameters was obtained from bootstrapping.\cite{bootstrap:1979}
Additionally, ASE-GUI and VMD were used to visualize the
trajectories.\cite{ase:2017,vmd:1996}\\

\section*{Supporting Information} 
The supporting material includes details of PhysNet, hyper-parameters,
validation on reaction pathways, lifetime distributions for
methyldioxirane, analysis for lifetimes of \textit{anti}-CI and
methyldioxirane, methyldioxirane concentration curves, additional
simulations at 1750 K and 2250 K, and data from additional training
including IRC structures around TS7.\\

\section*{Data Availability} 
The reference data that allow reproduction of the findings of this
study are openly available at
\url{https://github.com/MMunibas/anti-criegee}. \\

\section*{Acknowledgment}
This work has been financially supported by the Swiss National Science
Foundation (Grants 200021-117810, 200020-188724), and the University
of Basel (to MM).\\

\clearpage

\renewcommand{\thepage}{S\arabic{page}}
\renewcommand{\thetable}{S\arabic{table}}
\renewcommand{\thefigure}{S\arabic{figure}}
\renewcommand{\theequation}{S\arabic{equation}}
\renewcommand{\thesection}{S\arabic{section}} 
\setcounter{figure}{0}  
\setcounter{section}{0}  
\setcounter{table}{0}
\setcounter{page}{1}

\section*{Supporting Material}

\section{Details of PhysNet architecture}
PhysNet is a high-dimensional neural network (HDNN) of message-passing
type that uses the molecular configuration, i.e., the atomic numbers
and the Cartesian coordinates as inputs and energy, forces, charges,
and molecular dipole as outputs. \cite{MM.physnet:2019} In PhysNet,
learnable embeddings are assigned to each atom, and a Radial Basis
Function (RBF) with learnable parameters is applied to the distances
to obtain a representation of the molecule. Message passing is then
carried out between the atoms before passing the system to a stacked
set of modules. Each module contains a deep neural network (DNN) for
computing the intermolecular interactions as well as passing a part of
the results to the output block. Modules are also equipped with
residual layers to prevent overfitting. Finally, the output block
combines the results from all modules and computes the aforementioned
variables.  \\

\noindent
To train the model, a loss function involving a weighted sum of
energy, forces, total charge, and molecular dipole moment was used,
along with a non-hierarchical penalty to prevent overfitting; see
Equation \ref{sieqn:loss}. The model has been trained with the AMSGrad
optimizer. The hyperparameters used in PhysNet, which are not provided
in Section 2.1 of the paper, are tabulated in Table
\ref{sitbl:hyper}. Although PhysNet was originally implemented in
TensorFlow-1, a modified version involving TensorFlow-2 has been used
in this work. \cite{physnet-tf1, physnet-tf2} \\

\begin{eqnarray}
    \mathcal{L} &=& w_{E} \left| E - E^{ref} \right| + \frac{w_{F}}{3
      N} \sum_{i = 1}^{N} \sum_{\alpha = 1}^{3} \left| -
    \frac{\partial E}{\partial r_{i , \alpha}} - F_{i , \alpha}^{ref}
    \right| \nonumber \\ &+& w_{Q} \left| \sum_{i = 1}^{N} q_{i} -
    Q^{ref} \right| + \frac{w_{p}}{3} \sum_{\alpha = 1}^{3} \left|
    \sum_{i = 1}^{N} q_{i} r_{i , \alpha} - p_{\alpha}^{ref} \right| +
    \mathcal{L}_{nh}
    \label{sieqn:loss}
\end{eqnarray}
\clearpage

\section{Validation}
Starting from the \textit{anti-}CI intermediate, the
system first undergoes a rearrangement to methyldioxirane through TS1,
with a barrier of $\sim 19$ kcal mol$^{-1}$, followed by the
metastable intermediate IM1 through TS2. In the next step, the system
branches to CH$_4$ + CO$_2$, acetic acid, or methyl formate through
transition states TS3, TS4, or TS9, respectively. For all three
transition states, the energy barrier lies between 0.3 kcal mol$^{-1}$
and 2.6 kcal mol$^{-1}$.
\\

\noindent
From acetic acid, the system can form either of two products: 
CH$_4$ + CO$_2$, or formyl ketene (H$_2$C=C=O) + H$_2$O.
Acetic acid $\to$ CH$_4$ + CO$_2$ is an elementary reaction through
transition state TS5; see Figure \ref{fig:pes}B. On the other hand, 
Acetic acid $\to$ H$_2$C=C=O + H$_2$O can either take place as a 
single step, through TS8, or via the geminal diol intermediate, as 
shown in Figure \ref{fig:pes}C. Acetic acid $\to$ geminal diol $\to$
H$_2$C=C=O + H$_2$O reaction takes place through transition states
TS6 and TS7, respectively.
\\

\noindent
The branching to methyl formate from IM1 further leads to
fragmentation products methanol (CH$_3$OH) + CO. This elementary step
takes place through TS9, with a barrier of 69.4 kcal mol$^{-1}$; see
Figure \ref{fig:pes}D.

\clearpage

\section{Supporting Figures}

\begin{figure}[!hbtp]
    \centering

    \subfloat[Criegee $\rightarrow$ Methyldioxirane $\rightarrow$ CH$_4$ + CO$_2$]{%
        \includegraphics[height=2.5cm]{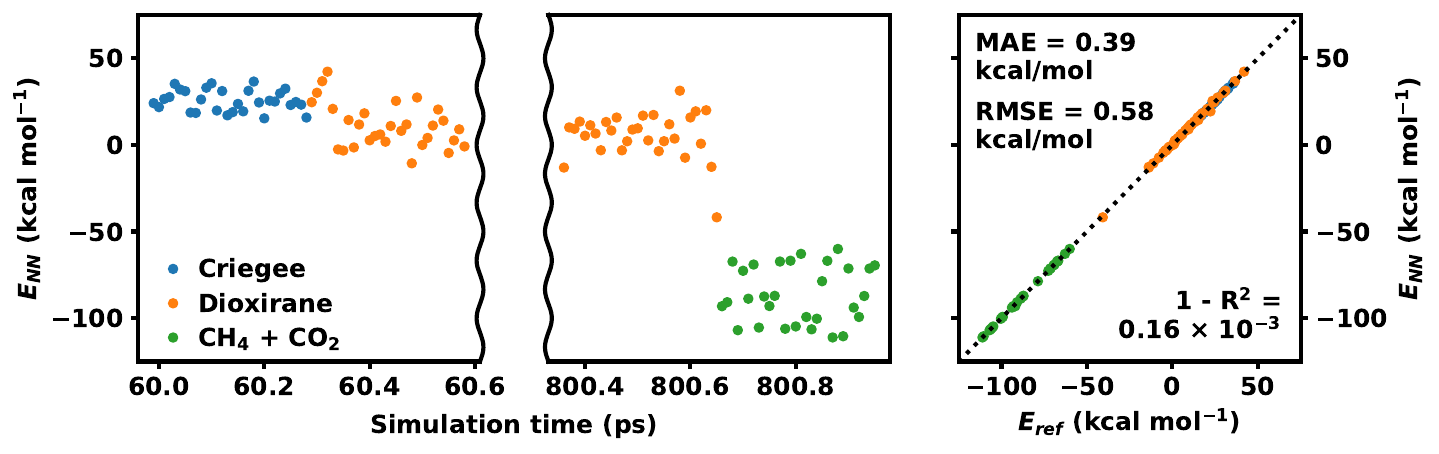}
    }

    \subfloat[Criegee $\rightarrow$ Methyldioxirane $\rightarrow$ Acetic acid $\rightarrow$ $\rightarrow$ \ce{H2C=C=O} + \ce{H2O}]{%
        \includegraphics[height=2.5cm]{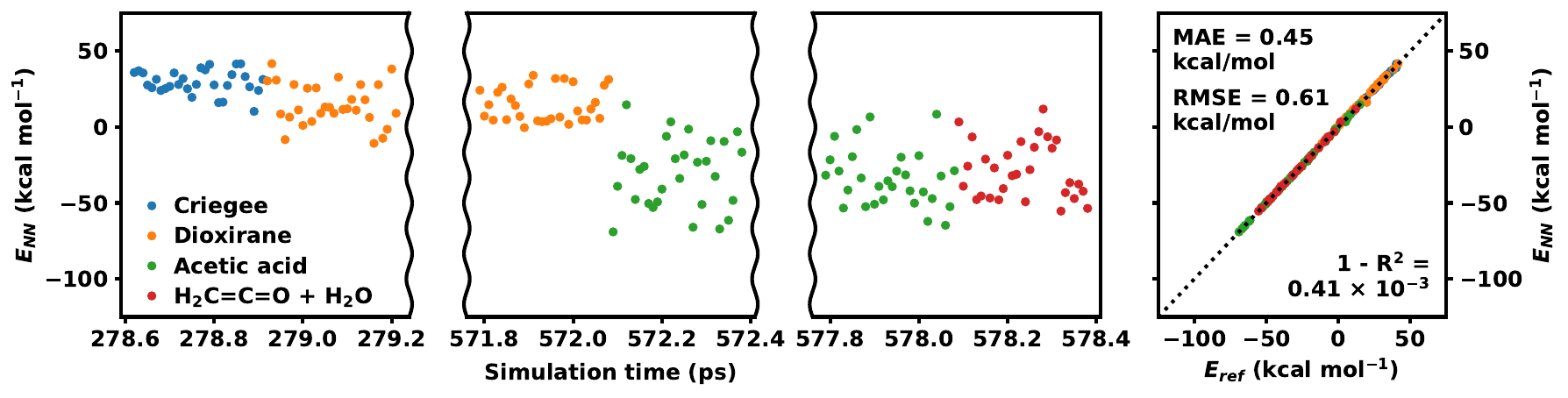}
    }

    \subfloat[Criegee $\rightarrow$ Methyldioxirane $\rightarrow$ Acetic acid $\rightarrow$ CH$_4$ + CO$_2$]{%
        \includegraphics[height=2.5cm]{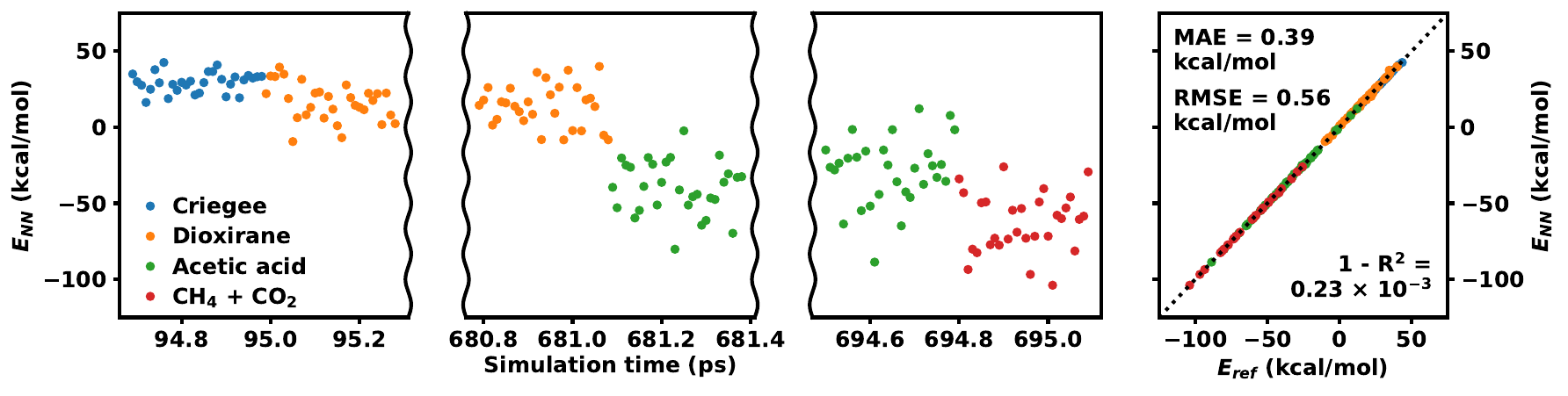}
    }

    \subfloat[Criegee $\rightarrow$ Methyldioxirane $\rightarrow$ Acetic acid $\rightarrow$ Geminal diol $\rightarrow$ \ce{H2C=C=O} + \ce{H2O}]{%
        \includegraphics[height=2.5cm]{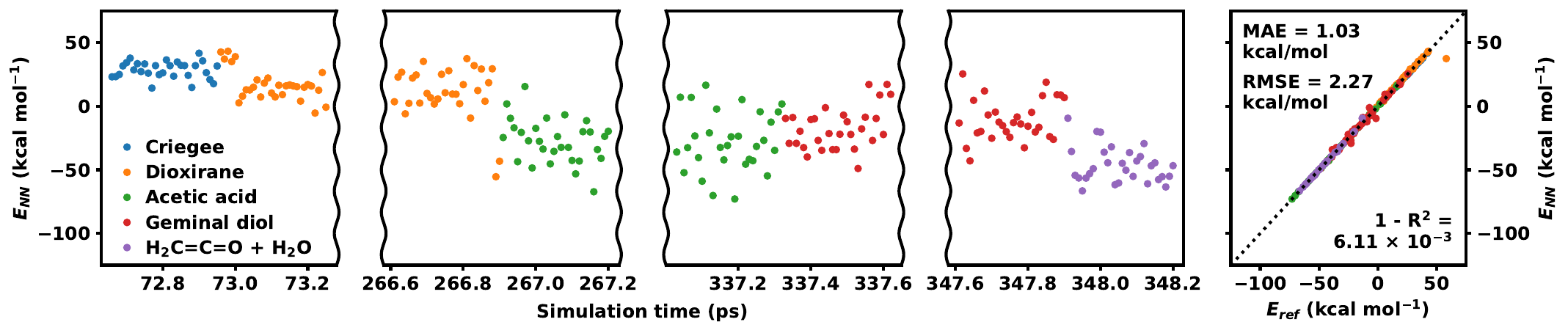}
    }

    \subfloat[Criegee $\rightarrow$ Methyldioxirane $\rightarrow$ Methyl formate $\rightarrow$ CH$_3$OH + CO]{%
        \includegraphics[height=2.5cm]{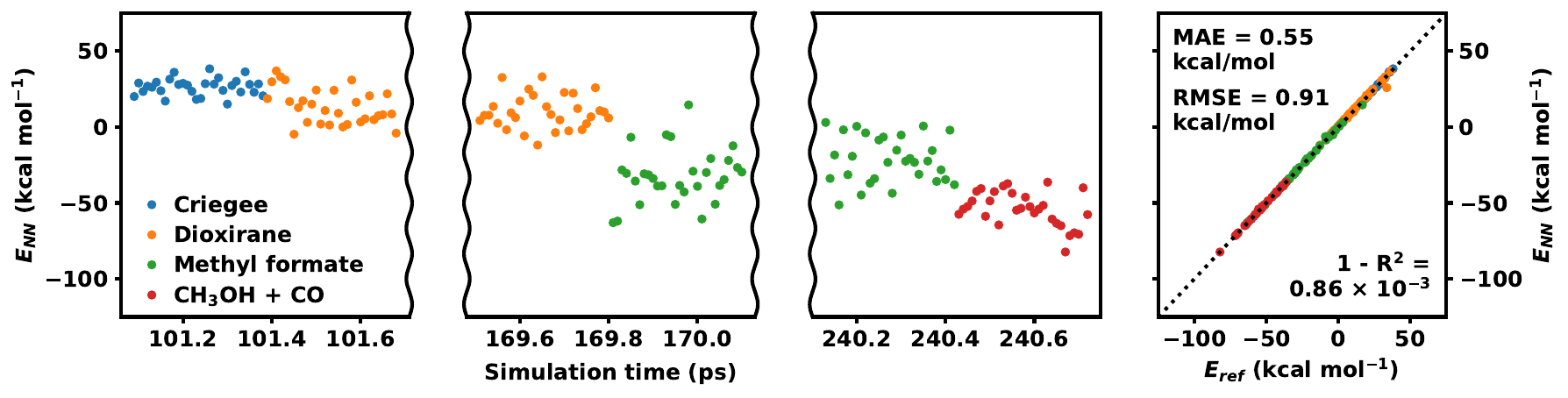}
    }
    
    \caption{Validation of the PhysNet model for various reaction
      pathways. All panels except the rightmost show 0.6 ps around
      transitions between reactant, intermediate(s) and/or
      products. The rightmost panel shows reference (MP2/aug-cc-pVTZ)
      vs. PhysNet energy for the structures near the
      transition. Transition states and IM1 are omitted due to their
      short lifetimes.}
    \label{sifig:valid-traj}
\end{figure}

\begin{figure}
    \centering
    \includegraphics[width=0.7\linewidth]{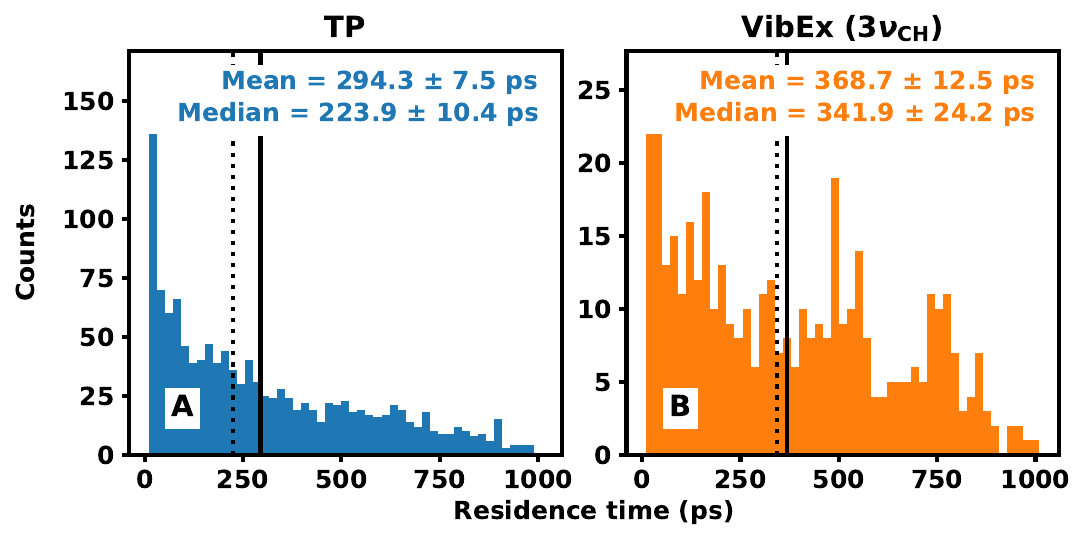}
    \caption{Residence time distribution of methyldioxirane in all
      trajectories that decompose to fragmentation products for TP
      (panel A) and VibEx (panel B) with 3 quanta of energy. The solid
      and dotted black vertical lines indicate the mean and the
      median, respectively.}
    \label{sifig:res-time-dioxi}
\end{figure}

\begin{figure}
    \centering
    \includegraphics[width=\linewidth]{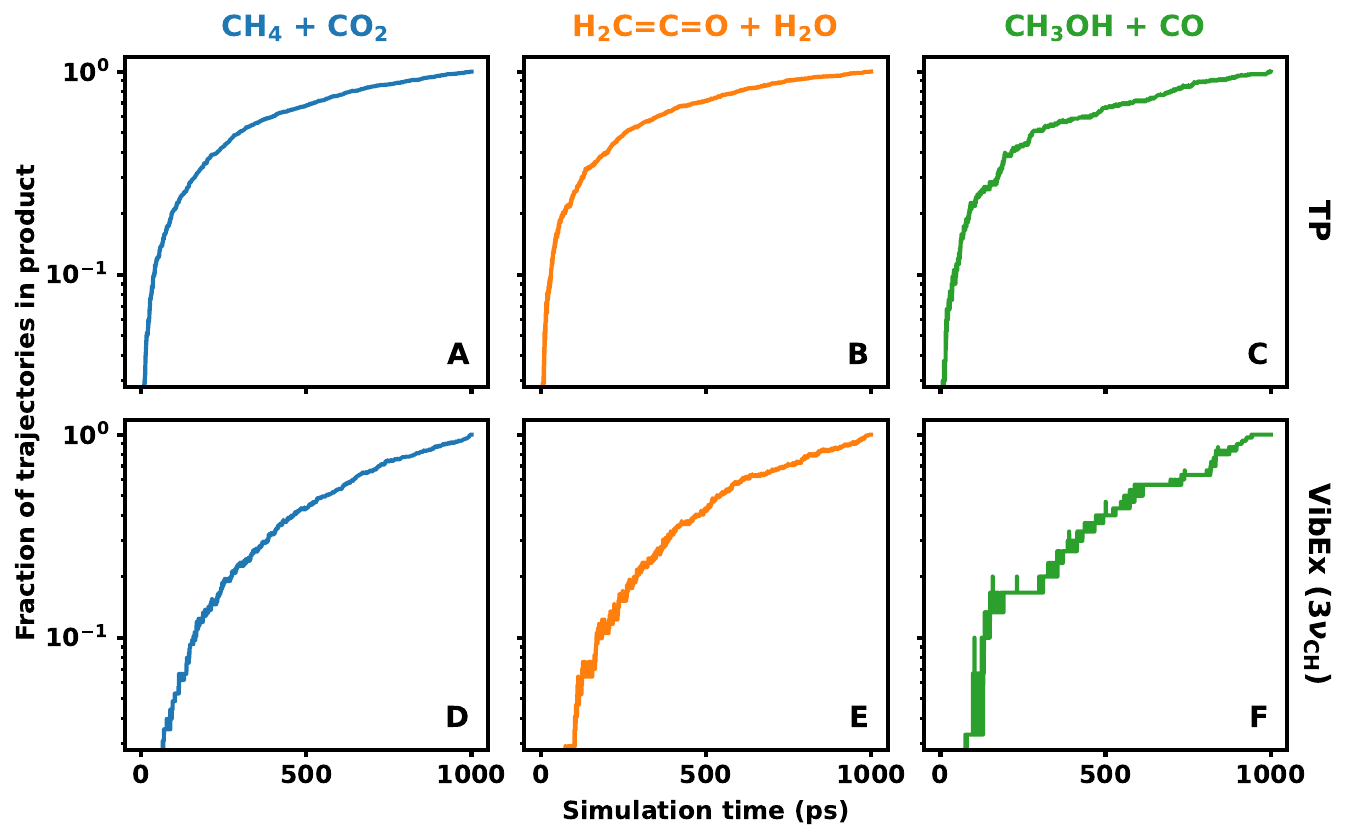}
    \caption{Fraction of the products (log scale) formed as a function
      of simulation time for TP (panels A-C) and VibEx preparation
      (panels D-F, 3 quanta). Products formed include CH$_4$ + CO$_2$
      (panels A/D), \ce{H2C=C=O} + H$_2$O (panels B/E), and CH$_3$OH +
      CO (panels C/F).}
    \label{sifig:product-log}
\end{figure}

\begin{figure}
    \centering
    \includegraphics[width=\linewidth]{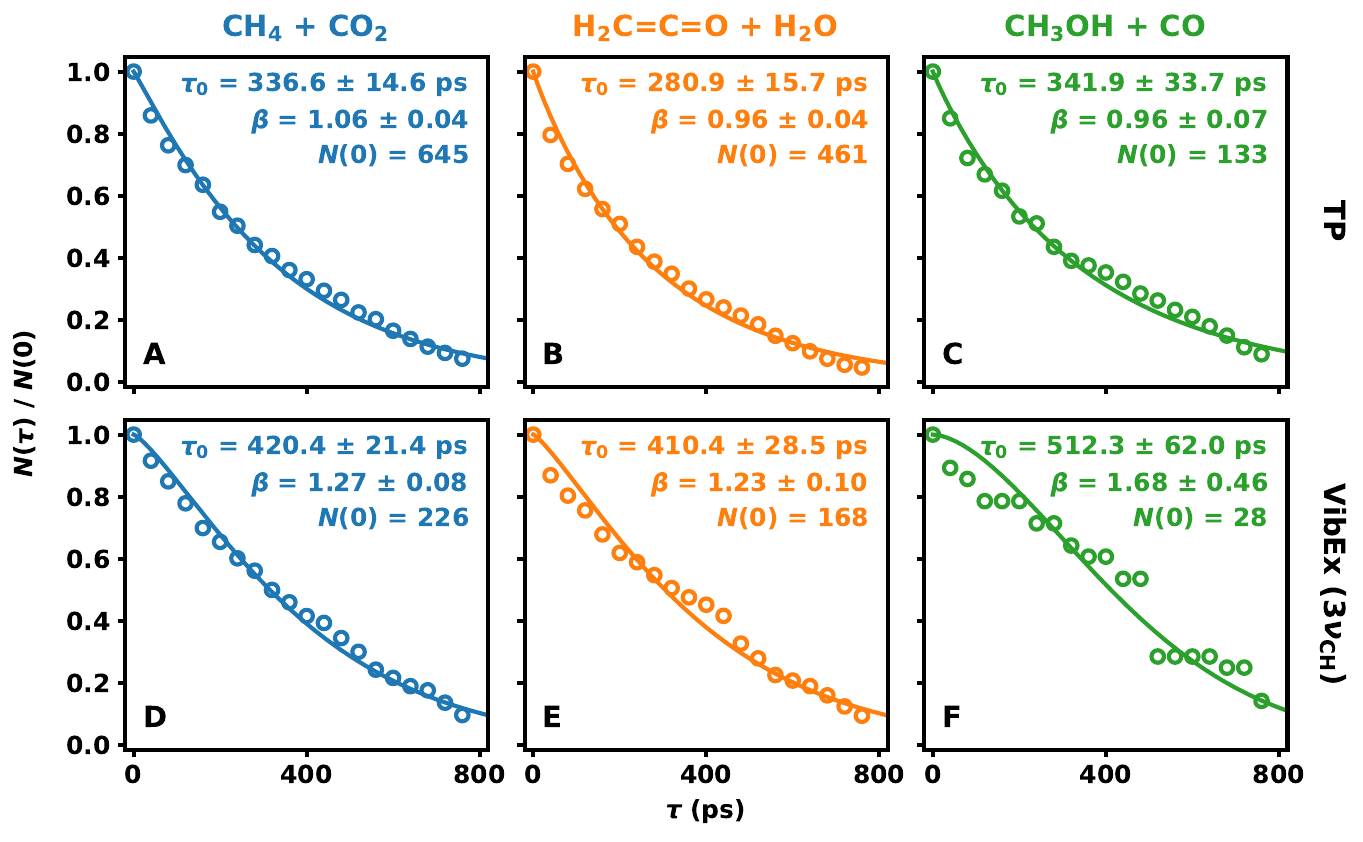}
    \caption{Survival probability of methyldioxirane for trajectories
      in a product channel for TP (panels A-C) and VibEx with 3 quanta
      of energy (panels D-F). Panels A, D: CH$_4$ + CO$_2$, Panels B,
      E: \ce{H2C=C=O} + H$_2$O, and Panels C, F: CH$_3$OH + CO. The
      fitting parameters for the stretched exponential fit function
      $\left[N(\tau)/N(0)=\exp\left(-(\tau/\tau_0)^\beta\right)\right]$
      are provided, along with $N(0)$, i.e., the total number of
      trajectories involved in the transition. The errors are computed
      using bootstrapping.\cite{bootstrap:1979}}
      \label{sifig:product-dioxi}
\end{figure}

\begin{figure}
    \centering
    \includegraphics[width=\linewidth]{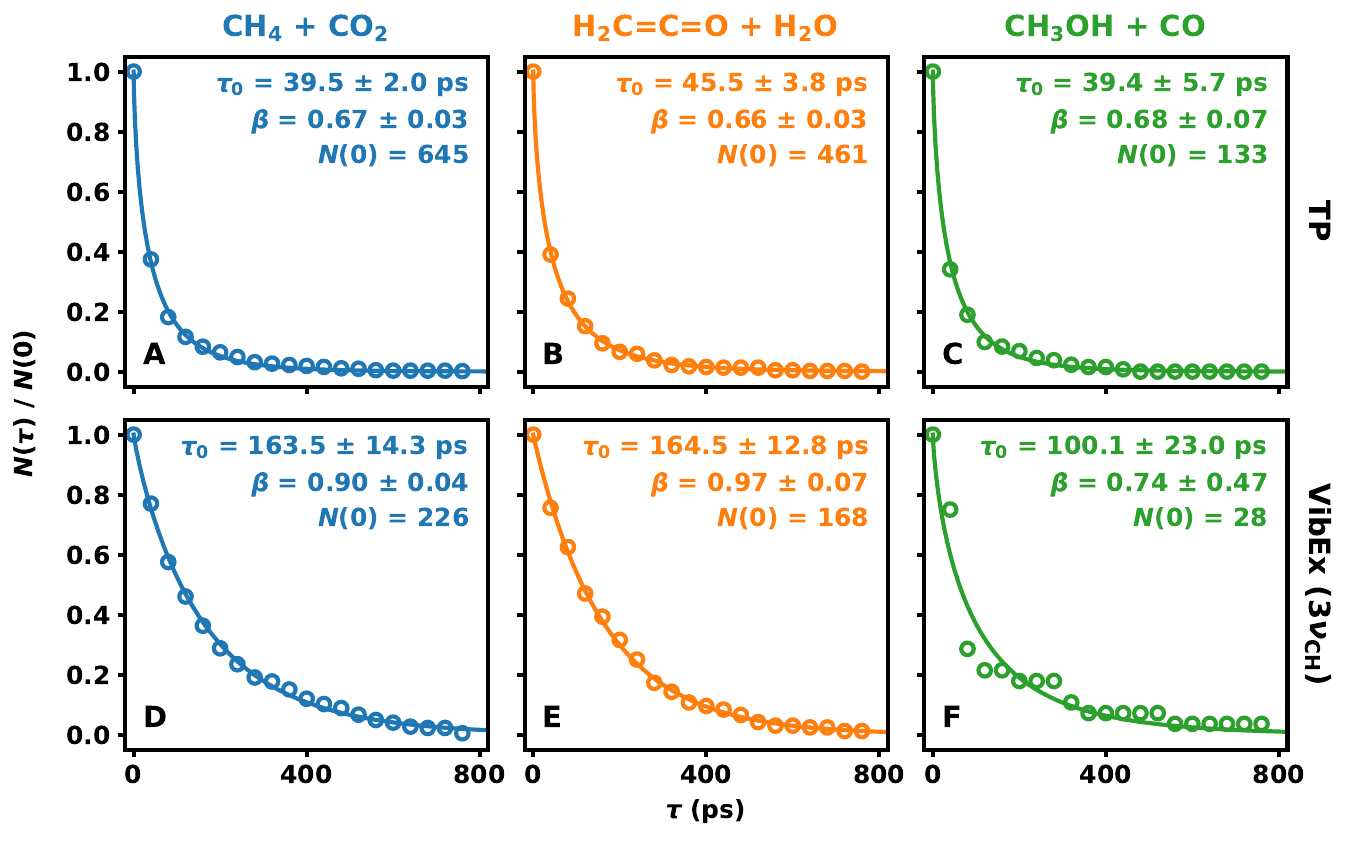}
    \caption{Survival probability of \textit{anti}-CI in for
      trajectories in a product channel for TP (panels A-C) and VibEx
      with 3 quanta of energy (panels D-F). Panels A, D: CH$_4$ +
      CO$_2$, Panels B, E: \ce{H2C=C=O} + H$_2$O, and Panels C, F:
      CH$_3$OH + CO. The fitting parameters for the stretched
      exponential fit function
      $\left[N(\tau)/N(0)=\exp\left(-(\tau/\tau_0)^\beta\right)\right]$
      are provided, along with $N(0)$, i.e., the total number of
      trajectories involved in the transition. The errors are computed
      using bootstrapping.\cite{bootstrap:1979}}
      \label{sifig:product-criegee}
\end{figure}
 
\begin{figure}
    \centering
    \includegraphics[width=\linewidth]{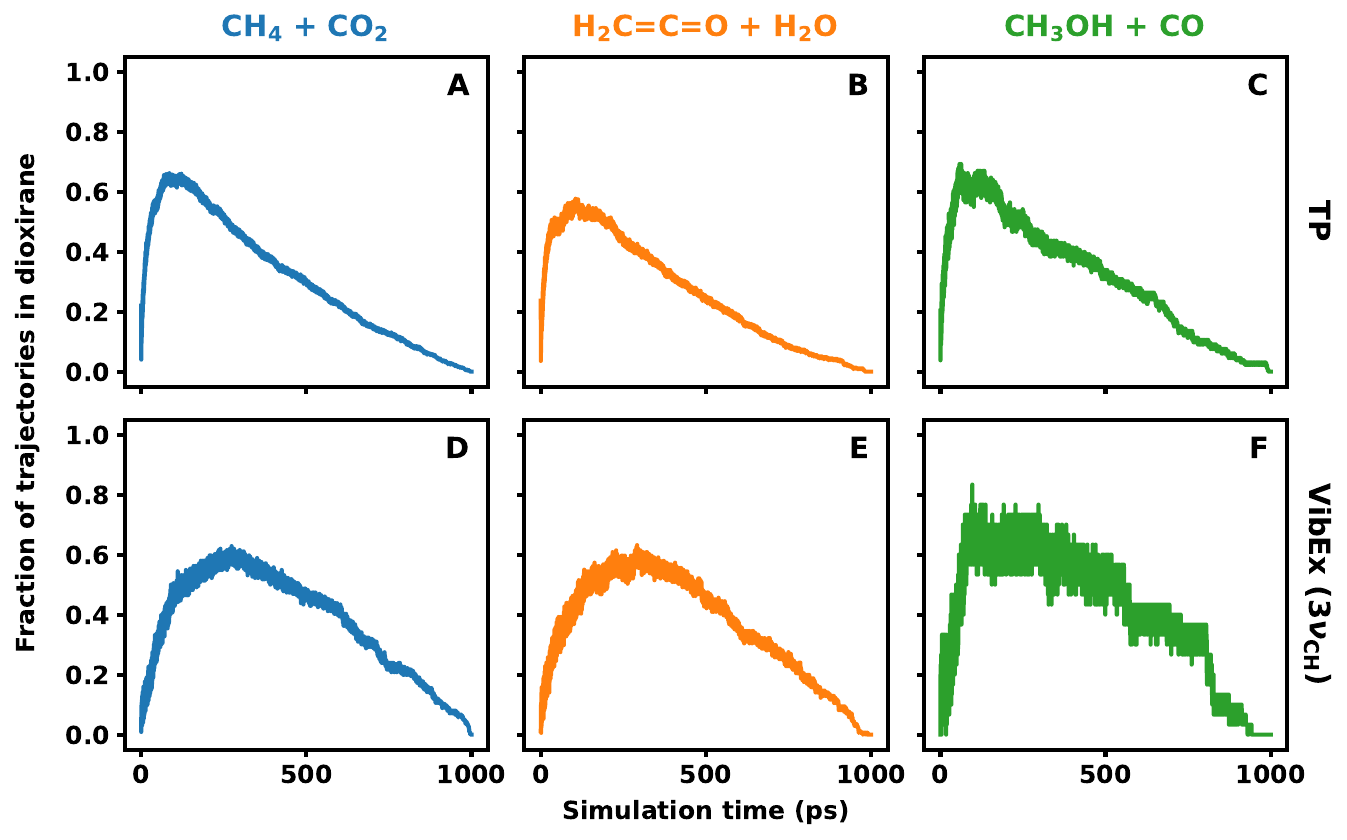}
    \caption{Fraction of the methyldioxirane with simulation time for
      TP (panels A-C) and vibrationally excited pathway with 3 quanta
      of energy (panels D-F). Panels A, D: CH$_4$ + CO$_2$, Panels B,
      E: \ce{H2C=C=O} + H$_2$O, and Panels C, F: CH$_3$OH + CO.}
    \label{sifig:dioxi-fraction}
\end{figure}

\clearpage
\section{Additional Figures for validation - other internal energies}

\begin{figure}
    \centering
    \includegraphics[width=0.6\linewidth]{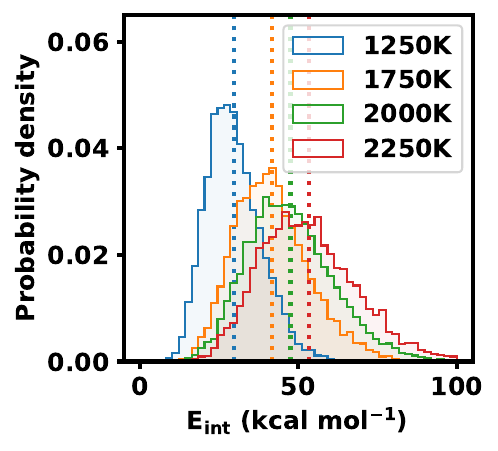}
    \caption{Probability distribution of internal energy $E_{\rm int}$
      at 1250 K, 1750 K, 2000 K and 2250 K. The dotted line
      corresponds to average internal energy, $\langle E_{\rm
        int}\rangle \sim \frac{3}{2}k_B T$, in accord with the
      equipartition theorem.}
    \label{sifig:int-e}
\end{figure}

\begin{figure}[h!]
  \centering
  \includegraphics[width=0.99\linewidth]{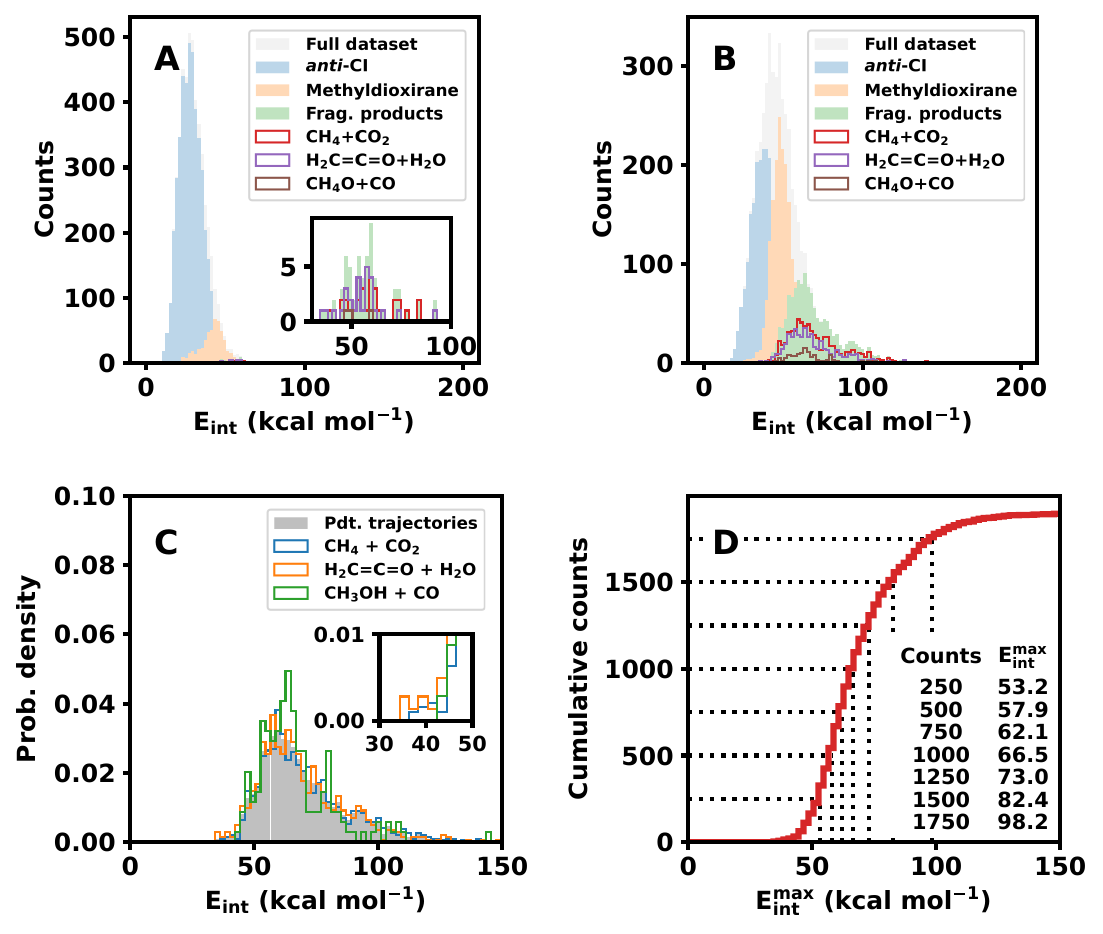}
  \caption{Initial distribution of energy for various trajectory end
    states and products. Panel A: $\langle E_{\rm int}\rangle
    \sim 29.3$ kcal mol$^{-1}$, Panel B: $\langle E_{\rm
      int}\rangle \sim 47.5$ kcal mol$^{-1}$, Panel C: All product-forming
      trajectories. Inset: Minimum energy required for each channel.
      Panel D. Cumulative distribution of all product-forming
    trajectories. Inset: Table with minimum energy requirement for
    every 250 product-forming trajectories. }
  \label{sifig:eint}
\end{figure}

\begin{figure}[h!]
  \centering
  \includegraphics[width=0.4\linewidth]{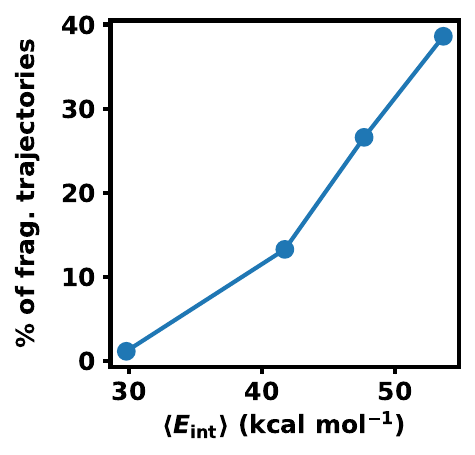}
  \caption{Percentage of fragmentation trajectories versus $\langle
    E_{\rm int}\rangle$.}
    \label{sifig:frag-prob}  
\end{figure}

\begin{figure}
    \centering
    \includegraphics[width=0.7\linewidth]{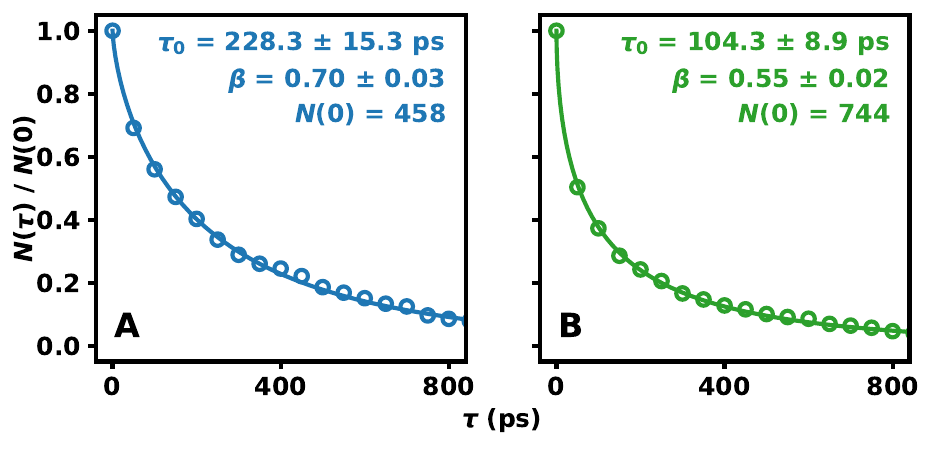}
    \caption{Survival probability $N(\tau)/N(0)$ of Criegee in all
      trajectories that proceed over TS1 for TP with $\langle E_{\rm
        int}\rangle \sim 41.8$ kcal mol$^{-1}$ and B. $\langle E_{\rm
        int}\rangle \sim 53.5$ kcal mol$^{-1}$). Clearly, the kinetics
      is inversely proportional to the energy in the system. In all
      panels, raw data and the fit to the stretched exponential
      function
      $\left[N(\tau)/N(0)=\exp\left(-(\tau/\tau_0)^\beta\right)\right]$
      are shown in the open circles and solid lines, respectively. The
      fitting parameters $\tau_0$ and $\beta$ are provided, along with
      $N(0)$, i.e., the total number of trajectories involved in the
      transition. The kinetics is inversely proportional to the energy
      in the system.  The errors in the fit are computed using
      bootstrapping.\cite{bootstrap:1979}}
    \label{sifig:criegee-survival}
\end{figure}

\begin{figure}
    \centering
    \includegraphics[width=\linewidth]{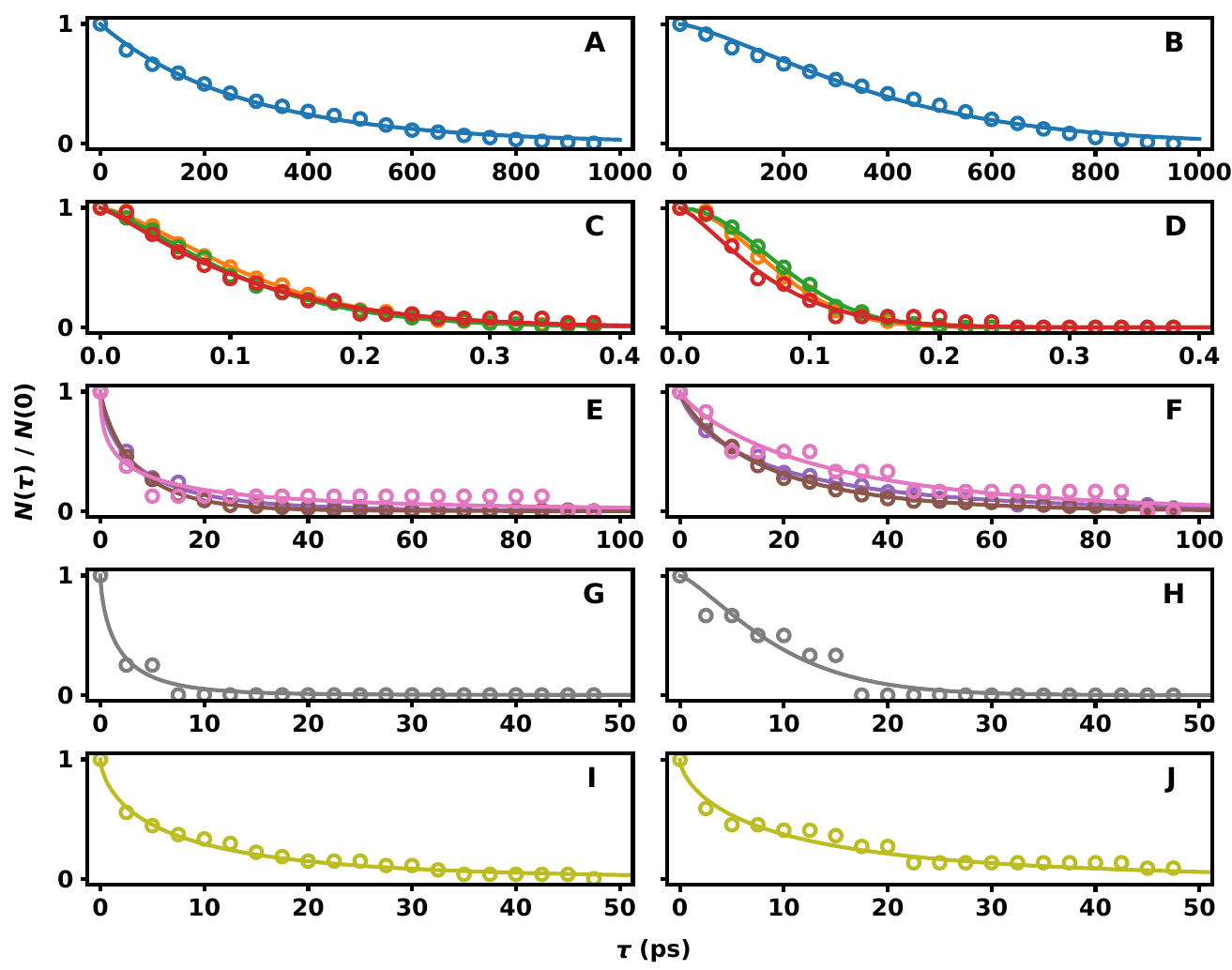}
    \caption{Survival probabilities for various intermediates
      corresponding to individual reaction channels for TP with
      $\langle E_{\rm int}\rangle \sim 53.5$ kcal mol$^{-1}$ (left) and
      $35 <E_{\rm int}< 53.2$ kcal mol$^{-1}$ (right). Panel
      A,B: Methyldioxirane $\rightarrow$ IM1 (blue).  Panel C,D: IM1
      $\rightarrow$ [CH$_4$ + CO$_2$ (orange), acetic acid (green),
        methyl formate (red)]. Panel E,F: Acetic acid $\rightarrow$
      [CH$_4$ + CO$_2$ (violet), \ce{H2C=C=O} + H$_2$O (brown),
        Geminal diol (magenta)]. Panels G,H: Geminal diol $\rightarrow$
      \ce{H2C=C=O} + H$_2$O (gray). Panel I,J: Methyl formate
      $\rightarrow$ CH$_3$OH + CO (lime). In all panels, the open
      circles and the solid lines correspond to the raw data and the
      fit to a stretched exponential function
      $\left[N(\tau)/N(0)=\exp\left(-(\tau/\tau_0)^\beta\right)\right]$. The
      parameters from the fit are summarized in Table
      \ref{sitbl:frag-kin}.}
    \label{sifig:frag-kin}
\end{figure}

\clearpage

\section{Figures for PhysNet Training}

\begin{figure}[h!]
    \centering
    \includegraphics[width=\linewidth]{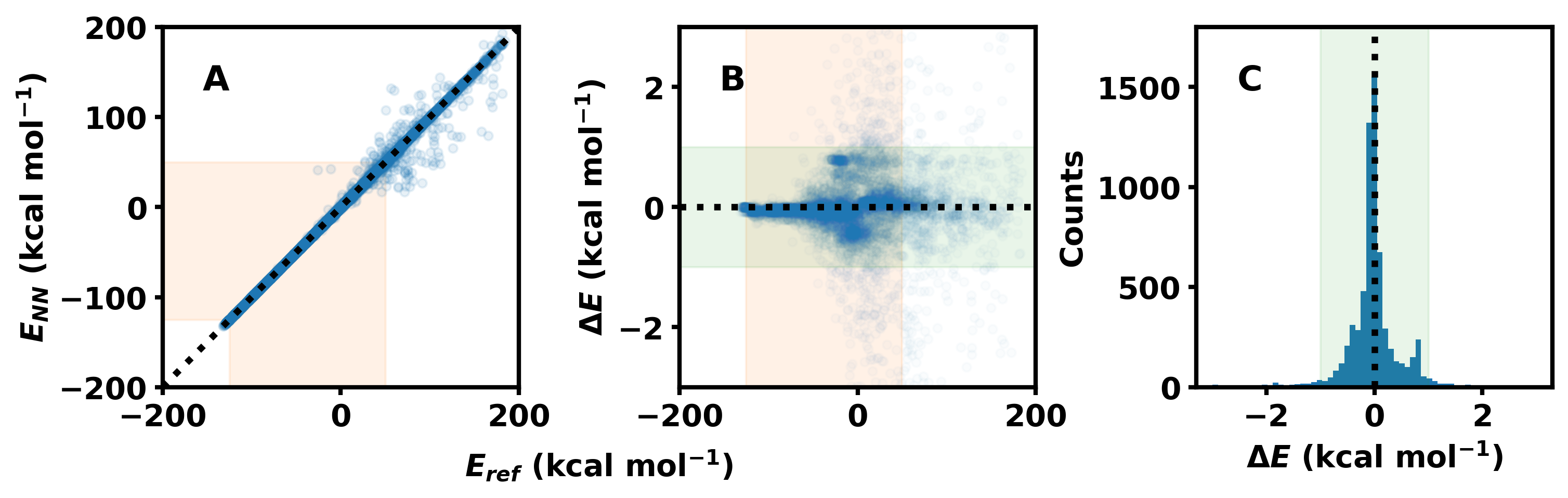}
    \caption{Validation of model PhysNet-4 on a test set ($\sim 7000$
      structures). Panel A: Reference energy (from MP2/aug-cc-pVTZ)
      against the PhysNet energy. Panel B: Error of energy ($\Delta E
      = E_{\text{PhysNet}}-E_{\text{MP2/aug-cc-pVTZ}}$) for the
      dataset along the reference energy. Panel C: Histogram of
      $\Delta E$ across the test dataset. The orange region indicates
      the energy range covered by the MD simulations, whereas the
      green region indicates an error of $\pm 1$ kcal mol$^{-1}$. For
      PhysNet-4, ${\rm MAE}(E) = 1.04$ kcal mol$^{-1}$ and correlation
      coefficient $R^2 \approx 0.993$.}
    \label{sifig:valid-test}
\end{figure}

\clearpage

\section{Figure for multi-reference effects}

\begin{figure}[h!]
    \centering
    \includegraphics[width=0.6\linewidth]{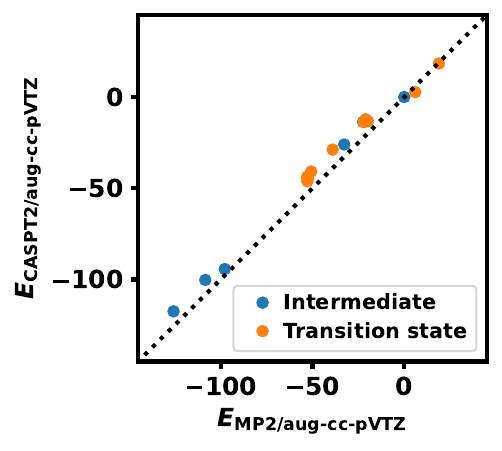}
    \caption{CASPT2 energy vs. MP2 energy for all intermediates (blue)
      and transition states (orange). The black dotted line indicates
      $E_\text{MP2}=E_\text{CASPT2}$.}
    \label{sifig:mr}
\end{figure}
\clearpage

\section{Table for final states}

\begin{table}[h!]
    \centering
    \begin{tabular}{l|cccc|cc}
        \hline
         Preparation & \multicolumn{4}{c|}{TP} & \multicolumn{2}{c}{VibEx}\\
         Temperature / Quanta& 1250 K & 1750 K & 2000 K & 2250 K & 3$\nu_{\rm CH}$ & 2$\nu_{\rm CH}$ \\
         $\langle E_{\rm int}\rangle$ / $E_{\rm VibEx} $& 29.8 & 41.8 & 47.5 & 53.5 & 27.4 & 18.3 \\         
         \hline
         No. of trajectories & 5000 & 1000 & 5000 & 1000 & 5000 & 5000 \\
         \hline
         {\it anti-}CI        & $88.3 (0.5)$ & $54.2 (1.6)$ & $37.6 (0.7)$ & $25.6 (1.3)$ & $30.7 (0.7)$ & $81.8 (0.5)$ \\
         Methyldioxirane      & $10.3 (0.4)$ & $32.4 (1.5)$ & $35.3 (0.7)$ & $35.4 (1.6)$ & $60.3 (0.7)$ & $17.6 (0.6)$ \\
         Fragmentation        & $1.2  (0.2)$ & $13.3 (1.0)$ & $26.6 (0.6)$ & $38.6 (1.5)$ & $8.6  (0.4)$ & $0.5  (0.1)$ \\
         \hline
         
    \end{tabular}
    \caption{Final states of the trajectories in percentage, computed
      from MD simulations at various TP and VibEx. The errors (in
      parentheses) are computed using
      bootstrapping.\cite{bootstrap:1979} Energies are reported in
      kcal mol$^{-1}$.}
    \label{sitbl:final-states}
\end{table}

\section{Table for fragmentation products}

\begin{table}[!hbtp]
    \centering
    \begin{tabular}{cc|ccc|ccc}
        \hline
         \multirow{2}{*}{Product} & \multirow{2}{*}{Precursor\ssc{a}} & \multicolumn{3}{c|}{TP with $\langle E_{\rm int}\rangle$ (kcal mol$^{-1}$)}
         & Mixed & Sub. & Addl. \\
         & &  29.8 & 41.8  & 53.5 & Dataset\ssc{b} & Sample\ssc{c} & Sample\ssc{d} \\
         
         \hline
         \multirow{3}{*}{\shortstack{\ce{CH4} \\+ \ce{CO2}}}
         & MD                 & $33.2 (5.9)$  & $33.7 (4.1)$  & $36.0 (2.4)$ & $31.5 (2.9)$ & $37.1 (2.7)$ & $38.3 (5.9)$ \\
         & AA                 & $14.4 (4.3)$  & $11.3 (2.7)$  & $14.0 (1.8)$ & $14.9 (2.3)$ & $11.4 (1.8)$ & $12.9 (3.9)$ \\
         & \textbf{All}       & $47.8 (6.5)$  & $45.0 (4.3)$  & $50.1 (2.5)$ & $46.3 (3.2)$ & $49.4 (2.8)$ & $51.4 (5.9)$ \\
         \hline
         \multirow{3}{*}{\shortstack{\ce{H2C=C=O} \\+ \ce{H2O}}}
         & AA                 & $39.8 (6.2)$ & $37.7 (4.3)$ & $32.4 (2.5)$ & $37.9 (2.8)$ & $32.8 (2.7)$ & $36.9 (5.9)$ \\
         & GD                 & $4.9  (2.7)$ & $2.3  (1.3)$ & $2.0  (0.7)$ & $2.5  (1.0)$ & $2.0  (0.8)$ & $2.9  (2.0)$ \\
         & \textbf{All}       & $44.3 (6.0)$ & $43.5 (4.3)$ & $35.8 (2.5)$ & $41.6 (3.2)$ & $36.2 (2.7)$ & $39.8 (5.8)$ \\
         \hline
         \multirow{2}{*}{\shortstack{\ce{CH3OH} \\+ \ce{CO}}}
         & MF                 & $4.7 (2.5)$ & $6.8 (2.1)$ & $7.0 (1.3)$ & $9.3 (1.8$) & $10.0 (1.7)$ & $7.2 (3.2)$\\
         & \textbf{All}       & $4.7 (2.5)$ & $6.8 (2.2)$ & $7.0 (1.3)$ & $9.3 (1.8$) & $10.0 (1.7)$ & $7.2 (3.2)$\\
         \hline
    \end{tabular}
    \\
    \ssc{a}: MD: Methyldioxirane, AA: Acetic acid, GD: Geminal diol, MF: Methyl formate.\\
    \ssc{b}: Mixed dataset from all samples, with $35 < E_{\rm int} < 53.5$ kcal mol$^{-1}$, 250 samples.\\
    \ssc{c}: Subsampling of 250 samples, repeated 1000 times, from TP with $\langle E_{\rm int}\rangle=47.5$ kcal mol$^{-1}$.\\
    \ssc{d}: Simulations with $35 < E_{\rm int} < 53.5$ kcal mol$^{-1}$, total 70 product-forming trajectories.\\

    \caption{Major fragmentation products, along with their immediate
      precursors, from MD simulations for TP and other sampling, in
      percentage.  The errors (in parentheses) are computed using
      bootstrapping.\cite{bootstrap:1979} Minor products and/or
      pathways (accessed by 20 or fewer trajectories) are not
      reported.}
    \label{sitbl:frag-products}
\end{table}

\clearpage
\section{Tables for retrained model}

\begin{table}[h!]
    \centering
    \begin{tabular}{c|cc}
        \hline
         \multirow{2}{*}{State} & \multicolumn{2}{c}{PhysNet model}\\
         & Original (PhysNet-4) & Retrained \\
         \hline
         {\it anti-}CI   & $25.6 \pm 1.3$ \%  & $25.4 \pm 1.4$ \% \\
         Methyldioxirane       & $35.4 \pm 1.6$ \%  & $34.9 \pm 1.5$ \% \\
         Fragmentation   & $38.6 \pm 1.5$ \%  & $39.2 \pm 1.5$ \% \\
         \hline
    \end{tabular}
    \caption{Comparison of original (PhysNet-4) and retrained models
      for the final states of the trajectories (1000 each), computed
      from MD simulations for TP with $E_{\rm int} = 53.5$ kcal
      mol$^{-1}$. The errors are computed using
      bootstrapping.\cite{bootstrap:1979}}
    \label{sitbl:retrain-final-states}
\end{table}

\begin{table}[h!]
    \centering
    \begin{tabular}{cc|cc}
        \hline
         \multirow{2}{*}{Product} & \multirow{2}{*}{Precursor} & \multicolumn{2}{c}{PhysNet model}\\
         & & Original (PhysNet-4) & Retrained \\
         
         \hline
         \multirow{3}{*}{\ce{CH4} + \ce{CO2}}
         & Methyldioxirane/IM1 &       $36.0 \pm 2.4$ \% & $43.5 \pm 2.5$ \% \\
         & Acetic acid &         $14.0 \pm 1.8$ \% & $9.4  \pm 1.5$ \% \\
         & \textbf{Total} &      $50.1 \pm 2.5$ \% & $53.9 \pm 2.5$ \% \\
         \hline
         \multirow{3}{*}{\ce{H2C=C=O} + \ce{H2O}}
         & Acetic acid &     $32.4 \pm 2.5$ \% & $29.0 \pm 2.2$ \% \\
         & Geminal diol &    $2.0 \pm 0.7$ \%  & $3.1 \pm 0.8$ \%  \\
         & \textbf{Total} &  $35.8 \pm 2.5$ \% & $33.1 \pm 2.3$ \% \\
         \hline
         \multirow{2}{*}{\ce{CH3OH} + \ce{CO}}
         & Methyl formate & $7.0 \pm 1.3$ \% & $8.2 \pm 1.4$ \% \\
         & \textbf{Total} & $7.0 \pm 1.3$ \% & $8.4 \pm 1.4$ \% \\
         \hline
    \end{tabular}
    \caption{Comparison of original (PhysNet-4) and retrained models,
      for major fragmentation products, along with their immediate
      precursors, from MD simulations for TP with $\langle E_{\rm
        int}\rangle \sim 53.5$ kcal mol$^{-1}$). The errors are
      computed using bootstrapping.\cite{bootstrap:1979}}
    \label{sitbl:retrain-frag-products}
\end{table}

\begin{table}[h!]
    \centering
    \begin{tabular}{ll|cc|cc}
        \hline
         \multirow{3}{*}{Initial} & \multirow{3}{*}{Final} & \multicolumn{4}{c}{PhysNet model}
         \\
         && \multicolumn{2}{c|}{Original (PhysNet-4)} & \multicolumn{2}{c}{Retrained}
         \\
         \cline{3-6}
         & & $\tau_0$ (ps) & $\beta$ & $\tau_0$ (ps) & $\beta$ \\
         \hline
         Methyldioxirane & IM1 & $278.11 \pm 17.9$  & $0.98 \pm 0.0$ & $270.27 \pm 17.8$ & $0.93 \pm 0.0$ \\
         \hline
         \multirow{3}{*}{IM1} 
         & CH$_4$ + CO$_2$ & $0.13 \pm 0.0$ & $1.42 \pm 0.1$ & $0.12 \pm 0.0$ & $1.53 \pm 0.1$\\
         & Acetic acid     & $0.12 \pm 0.0$ & $1.35 \pm 0.1$ & $0.13 \pm 0.0$ & $1.48 \pm 0.1$\\
         & Methyl formate  & $0.12 \pm 0.0$ & $1.19 \pm 0.2$ & $0.13 \pm 0.0$ & $1.23 \pm 0.3$\\
         \hline
         \multirow{3}{*}{Acetic acid} 
         & CH$_4$ + CO$_2$       & $7.23 \pm 1.5$ & $0.68 \pm 0.1$ & $10.50 \pm 1.8$ & $1.05 \pm 0.2$\\
         & \ce{H2C=C=O} + H$_2$O & $6.94 \pm 0.8$ & $0.84 \pm 0.1$ & $8.16 \pm 1.0$ & $0.80 \pm 0.1$\\
         & Geminal diol          & $5.87 \pm 6.9^\text{a}$ & $0.45 \pm 2.0^\text{a}$ & $6.43 \pm 2.9^\text{a}$ & $0.70 \pm 0.2^\text{a}$\\
         \hline
         Geminal diol & \ce{H2C=C=O} + H$_2$O & $1.93 \pm 1.4^\text{a}$ & $0.67 \pm 0.3^\text{a}$ & $3.41 \pm 1.2^\text{a}$ & $0.86 \pm 0.3^\text{a}$\\
         \hline
         Methyl formate & CH$_3$OH + CO & $7.18 \pm 3.0$ & $0.64 \pm 0.1$ &  $6.84 \pm 1.6$ & $0.76 \pm 0.2$ \\
         \hline
    \end{tabular}
    \\
    $^\text{a}$: Unreliable error estimate due to lack of data
    \caption{Comparison of original (PhysNet-4) and retrained models
      on the stretched exponential fit to the survival probabilities,
      i.e. $\left[N(\tau)/N(0)=\exp\left(-(\tau/\tau_0)^\beta\right)\right]$,
      for different species in each reaction channel, for TP with
      $\langle E_{\rm int}\rangle \sim 53.5$ kcal
      mol$^{-1}$). $\tau_0$ and $\beta$ correspond to the
      characteristic timescale and stretching exponent,
      respectively. The errors are computed with
      bootstrapping.\cite{bootstrap:1979}}
    \label{sitbl:retrain-frag-kin}
\end{table}

\clearpage
\section{Table for stretched exponential fits}

\begin{table}
    \centering
    \begin{tabular}{ll|cc|cc|cc}
      \hline
      \multirow{4}{*}{Initial\ssc{a}} & \multirow{4}{*}{Final\ssc{a}} & \multicolumn{2}{c|}{TP with} & \multicolumn{2}{c|}{Mixed dataset} & \multicolumn{2}{c}{Subsampling at }
         \\
         && \multicolumn{2}{c|}{$\langle E_{\rm int} \rangle \sim 53.5$ }& \multicolumn{2}{c|}{$35 < E_{\rm int} < 53.2$}& \multicolumn{2}{c}{$\langle E_{\rm int} \rangle \sim 47.5$ }
         \\
         && \multicolumn{2}{c|}{kcal mol$^{-1}$}&\multicolumn{2}{c|}{kcal mol$^{-1}$}& \multicolumn{2}{c}{kcal mol$^{-1}$}
         \\
         \cline{3-8}
         & & $\tau_0$ (ps) & $\beta$ & $\tau_0$ (ps) & $\beta$ & $\tau_0$ (ps) & $\beta$ \\
         \hline
         MD & IM1 & $278.1 (17.9)$  & $1.0 (0.0)$  & $418.5 (24.0)$ & $1.4 (0.1)$ & $318.3 (42.4) $ & $1.0(0.1)$\\
         \hline
         \multirow{3}{*}{IM1} 
         & P1 & $0.1 (0.0)$ & $1.4 (0.1)$ & $0.1 (0.0)$ & $1.9 (0.2)$ & $0.1(0.0)$ & $1.3 (0.2)$\\
         & AA     & $0.1 (0.0)$ & $1.4 (0.1)$ & $0.1 (0.0)$ & $2.0 (0.2)$& $0.1(0.0)$ & $1.4 (0.3)$\\
         & MF  & $0.1 (0.0)$ & $1.2 (0.2)$ & $0.1 (0.0)$ & $1.3 (0.4)$& $0.1(0.0)$ & $1.6 (0.7)$\\
         \hline
         \multirow{3}{*}{AA} 
         & P1       & $7.2 (1.5)$ & $0.7 (0.1)$ & $18.2 (5.2)$ & $0.7 (0.1)$ & $9.5(5.0)$ & $0.8 (0.4)$\\
         & P2 & $7.0 (0.8)$ & $0.8 (0.1)$ & $16.6 (1.9)$ & $0.9 (0.1)$ & $9.1(2.9)$ & $0.7 (0.2)$\\
         & GD          &\multicolumn{2}{c|}{Insufficient data} & \multicolumn{2}{c|}{Insufficient data} & \multicolumn{2}{c}{Insufficient data} \\
         \hline
         GD & P2 &\multicolumn{2}{c|}{Insufficient data} & \multicolumn{2}{c|}{Insufficient data} & \multicolumn{2}{c}{Insufficient data}\\
         \hline
         MF & P3 & $7.2 (3.0)$ & $0.6 (0.1)$ & $10.2 (4.8)$ & $0.7 (0.2)$ & $4.9 (3.7)$ & $0.6 (0.3)$\\
         \hline
    \end{tabular}
    \\
    \ssc{a}: MD: Methyldioxirane, AA: Acetic acid, GD: Geminal diol, MF: Methyl formate, \\
    P1: CH$_4$ + CO$_2$, P1: \ce{H2C=C=O} + H$_2$O, P3: CH$_3$OH + CO.
    \caption{Stretched exponential fit to the survival probabilities,
      i.e. $\left[N(\tau)/N(0)=\exp\left(-(\tau/\tau_0)^\beta\right)\right]$,
      for different species in each reaction channel, for the TP with
      $\langle E_{\rm int}\rangle \sim 53.5$ kcal mol$^{-1}$, mixed
      dataset with $35 < E_{\rm int} < 53.2$ kcal mol$^{-1}$ and
      subsampling from TP with $\langle E_{\rm int}\rangle \sim 47.5$
      kcal mol$^{-1}$. $\tau_0$ and $\beta$ correspond to the
      characteristic timescale and stretching exponent,
      respectively. The errors (in parentheses) are computed with
      bootstrapping.\cite{bootstrap:1979}}
    \label{sitbl:frag-kin}
\end{table}

\clearpage
\section{Tables for Training PhysNet}

\begin{table}[!hbtp]
    \begin{tabular}{ccc}
        \hline
        Hyper-parameter & Description & Value\\
        \hline
        $F$ & Feature-space dimensionality & 128\\
        $K$ & Number of RBF & 64 \\
        $N_\text{module}$ & Number of modules & 5 \\
        $N_\text{residual}^\text{atomic}$ & Number of residual blocks (atomic) & 3\\
        $N_\text{residual}^\text{interaction}$ & Number of residual blocks (interaction) & 2\\
        $N_\text{residual}^\text{output}$ & Number of residual blocks (output) & 1\\
        $r_\text{cutoff}$ & cut-off radius in interaction block & 10 \AA\\
        & Random number seed & 53, 54, 55, 56\\
        & Batch size (training) & 64 \\
        & Batch size (validation) & 60 \\
        & Validation interval & 100 \\
        & Save interval & 100 \\
        & Electrostatics & Yes \\
        & Dispersion & No \\
        \hline
    \end{tabular}
    \caption {Hyperparameters used in PhysNet training}
    \label{sitbl:hyper}
\end{table}

\begin{table}
\caption{Validation of PhysNet models on the full test set ($\sim
  7000$ structures) as well as the subset with $E \in [-135,35]$ kcal
  mol$^{-1}$, i.e., those relevant to MD simulation. The highlighted
  model was used for MD simulations, owing to the lowest MAE and RMSE
  over the full test set. Note that the forces reported here are used
  for training and validation, while for MD simulations, they are
  directly computed using the energies. PhysNet-4 was chosen based on
  the overall test set errors.} \centering
	\begin{tabular}{c|cc|cc|cc}
        \hline
		\multirow{4}{*}{Model} & \multicolumn{4}{c|}{Full test set} & \multicolumn{2}{c}{MD range}\\
        
        & \multicolumn{2}{c|}{Energy} & \multicolumn{2}{c|}{Forces} & \multicolumn{2}{c}{Energy} \\
        & \multicolumn{2}{c|}{(kcal mol$^{-1}$)} & \multicolumn{2}{c|}{(kcal mol$^{-1}$ \AA$^{-1}$)} & \multicolumn{2}{c}{(kcal mol$^{-1}$)}\\
		& MAE & RMSE & MAE & RMSE & MAE & RMSE \\
		\hline
		PhysNet-1 & 1.07 & 5.06 & 1.81 & 10.62 & 0.45 & 1.72 \\
		PhysNet-2 & 1.04 & 5.60 & 1.81 & 18.28 & 0.41 & 1.31 \\
		PhysNet-3 & 1.20 & 5.90 & 1.91 & 18.32 & 0.48 & 1.89 \\
		\textbf{PhysNet-4} & 1.04 & 4.81 & 2.05 & 21.62 & 0.41 & 1.60 \\
		\hline
	\end{tabular}
    \label{sitbl:test-data}
\end{table}

\clearpage
\section{Table for multi-reference data}

\begin{table}[]
    \centering
    \begin{tabular}{lccc}
        \hline
        \multirow{2}{*}{Species} & \multicolumn{2}{c}{Energy (kcal mol$^{-1}$)} & T1-diagnostic\ssc{a}\\
        & MP2\ssc{b} & CASPT2\ssc{c} & ($\times 10^{-2}$)\\
        \hline
        {\it anti}-CI\ssc{d}    & 0.0    &  0.0    &  3.9  \\
        Methyldioxirane & -32.7  &  -26.0  &  1.5  \\
        IM1             & -22.4  &  -13.6  &  2.5  \\
        Acetic acid     & -125.7 &  -117.5 &  1.6  \\
        Geminal diol    & -97.8  &  -94.3  &  1.3  \\
        Methyl formate  & -108.4 &  -100.3 &  1.7  \\
        TS1             & 18.9   &  18.4   &  4.2  \\
        TS2             & 6.0    &  2.7    &  4.6  \\
        TS3             & -21.0  &  -12.2  &  2.4  \\
        TS4             & -22.1  &  -13.7  &  2.5  \\
        TS5             & -53.0  &  -43.9  &  1.9  \\
        TS6             & -52.3  &  -44.5  &  1.8  \\
        TS7             & -52.8  &  -46.1  &  1.6  \\
        TS8             & -50.7  &  -40.8  &  1.8  \\
        TS9             & -19.8  &  -13.3  &  2.9  \\
        TS10            & -39.0  &  -28.8  &  2.2  \\
        \hline
    \end{tabular}
    \\
    \ssc{a}: Computed with CCSD/aug-cc-pVDZ.\\
    \ssc{b}: Basis set: aug-cc-pVTZ.\\
    \ssc{c}: Basis set: aug-cc-pVTZ, active space: CAS(12,10).\\
    \ssc{d}: Used as reference (zero) for energy.
    \caption{Multi-reference effects in the intermediates and
      transition states using the MP2/aug-cc-pVTZ-optimized
      structures.}
    \label{sitab:multi-ref}
\end{table}
\clearpage

\bibliography{refs.clean}

@string{bj = "Biophys. J."}

@string{BJ = {Biophys. J.}}

@string{cj= "Comput. J."}

@string{csr = "Chem. Soc. Rev."}

@string{ESPI= "Environ. Sci.: Process. Impacts"}

@string{irpc = "Intern. Rev. Phys. Chem."}

@string{M = {Microbiology}}

@string{nature = "Nature"}

@string{pccp = "Phys. Chem. Chem. Phys."}

@string{PNAS = {Proc. Natl. Acad. Sci. USA}}

@string{pnas = "Proc. Nat. Acad. Sci. USA"}

@string{S = {Science}}

@string{sci = "Science"}

@string{St = {Structure}}

@string{structure = "Structure"}

@article{gordon88mp2,
	title        = {MP2 energy evaluation by direct methods},
	author       = {Martin Head-Gordan and Pople, John A. and Michael Frisch},
	year         = 1988,
	journal      = {Chem. Phys. Lett.},
	volume       = 153,
	number       = 6,
	pages        = 503
}

@Article{dunning89BtoNe,
  author  = {Dunning, Thom H.},
  title   = {Gaussian basis sets for use in correlated molecular calculations. I. The atoms boron through neon and hydrogen},
  journal = {J. Chem. Phys.},
  year    = {1989},
  volume  = {90},
  number  = {2},
  pages   = {1007--1023},
}

@misc{g16,
    author={M. J. Frisch and G. W. Trucks and H. B. Schlegel and G. E. Scuseria and M. A. Robb and J. R. Cheeseman and G. Scalmani and V. Barone and G. A. Petersson and H. Nakatsuji and X. Li and M. Caricato and A. V. Marenich and J. Bloino and B. G. Janesko and R. Gomperts and B. Mennucci and H. P. Hratchian and J. V. Ortiz and A. F. Izmaylov and J. L. Sonnenberg and D. Williams-Young and F. Ding and F. Lipparini and F. Egidi and J. Goings and B. Peng and A. Petrone and T. Henderson and D. Ranasinghe and V. G. Zakrzewski and J. Gao and N. Rega and G. Zheng and W. Liang and M. Hada and M. Ehara and K. Toyota and R. Fukuda and J. Hasegawa and M. Ishida and T. Nakajima and Y. Honda and O. Kitao and H. Nakai and T. Vreven and K. Throssell and Montgomery, {Jr.}, J. A. and J. E. Peralta and F. Ogliaro and M. J. Bearpark and J. J. Heyd and E. N. Brothers and K. N. Kudin and V. N. Staroverov and T. A. Keith and R. Kobayashi and J. Normand and K. Raghavachari and A. P. Rendell and J. C. Burant and S. S. Iyengar and J. Tomasi and M. Cossi and J. M. Millam and M. Klene and C. Adamo and R. Cammi and J. W. Ochterski and R. L. Martin and K. Morokuma and O. Farkas and J. B. Foresman and D. J. Fox},
    title={Gaussian˜16 {R}evision {C}.01},
    year={2016},
    note={Gaussian Inc. Wallingford CT}
}

@Article{MM.physnet:2019,
  author  = {Unke, Oliver T. and Meuwly, Markus},
  title   = {PhysNet: A Neural Network for Predicting Energies, Forces, Dipole Moments, and Partial Charges},
  journal = {J. Chem. Theory Comput.},
  year    = {2019},
  volume  = {15},
  number  = {6},
  pages   = {3678--3693},
}

@article{ase:2017,
  title={The atomic simulation environment -- a Python library for
                  working with atoms},
  author={Larsen, Ask Hjorth and Mortensen, Jens J{\o}rgen and
                  Blomqvist, Jakob and Castelli, Ivano E and
                  Christense n, Rune and Du{\l}ak, Marcin and Friis,
                  Jesper and Groves, Michael N and Hammer, Bj{\o}rk
                  and Hargus, Cory and other s},
  journal={J. Phys. Condens. Matter},
  volume={29},
  number={27},
  pages={273002},
  year={2017},
  publisher={IOP Publishing}
}

@article{khan:2018,
  title={Criegee intermediates and their impacts on the troposphere},
  author={Khan, MAH and Percival, CJ and Caravan, RL and Taatjes, CA
                  and Shallcross, DE},
  journal=ESPI,
  volume={20},
  number={3},
  pages={437--453},
  year={2018},
  publisher={Royal Society of Chemistry}
}

@Article{criegee:1949,
  author    = {Criegee, Rudolf and Wenner, Gotthilf},
  title     = {Die Ozonisierung des 9, 10-Oktalins},
  journal   = {Justus Liebigs Ann. Chem.},
  year      = {1949},
  volume    = {564},
  number    = {1},
  pages     = {9--15},
  publisher = {Wiley Online Library},
}

@Article{yin2017does,
  author    = {Yin, Cangtao and Takahashi, Kaito},
  title     = {How does substitution affect the unimolecular reaction rates of Criegee intermediates?},
  journal   = {Phys. Chem. Chem. Phys.},
  year      = {2017},
  volume    = {19},
  number    = {19},
  pages     = {12075--12084},
  publisher = {Royal Society of Chemistry},
}

@Article{lester:2016,
  author    = {Kidwell, Nathanael M and Li, Hongwei and Wang, Xiaohong and Bowman, Joel M and Lester, Marsha I},
  title     = {Unimolecular dissociation dynamics of vibrationally activated CH$_3$CHOO Criegee intermediates to OH radical products},
  journal   = {Nat. Chem.},
  year      = {2016},
  volume    = {8},
  number    = {5},
  pages     = {509--514},
  publisher = {Nature Publishing Group},
}

@Article{moller1934note,
  author    = {M{\o}ller, Chr and Plesset, Milton S},
  title     = {Note on an approximation treatment for many-electron systems},
  journal   = {Phys. Rev.},
  year      = {1934},
  volume    = {46},
  number    = {7},
  pages     = {618},
  publisher = {APS},
}

@Article{MM.atmos:2020,
  author  = {K{\"a}ser, Silvan and Unke, Oliver T and Meuwly, Markus},
  title   = {Isomerization and decomposition reactions of acetaldehyde relevant to atmospheric processes from dynamics simulations on neural network-based potential energy surfaces},
  journal = {J. Chem. Phys.},
  year    = {2020},
  volume  = {152},
  number  = {21},
  pages   = {214304},
}

@article{taatjes:2015,
  title={The physical chemistry of Criegee intermediates in the gas phase},
  author={Osborn, David L and Taatjes, Craig A},
  journal=irpc,
  volume={34},
  number={3},
  pages={309--360},
  year={2015},
  publisher={Taylor \& Francis}
}

@article{dyakov:2025,
  title={Dissociation Dynamics of CH$_3$CHOO Criegee Intermediates in
                  the Earth’s Atmosphere: Ab Initio and RRKM Study},
  author={Dyakov, Yuri A and Stepanov, Ilya G and Adamson, Sergey O
                  and Morozov, Igor I and Rodionov, Igor D and
                  Rodionova, Irina P and Shushin, Anatoly I and
                  Shestakov, Denis V and Teranishi, Yoshiaki and Wang,
                  Pao K and others},
  journal={ACS Earth Space Chem.},
  volume={9},
  number={3},
  pages={671--680},
  year={2025},
  publisher={ACS Publications}
}

@Article{wang:2019,
  author    = {Wang, Zhiping and Dyakov, Yuri A and Bu, Yuxiang},
  title     = {Dynamics Insight into Isomerization and Dissociation of Hot Criegee Intermediate CH$_3$CHOO},
  journal   = {J. Phys. Chem. A},
  year      = {2019},
  volume    = {123},
  number    = {5},
  pages     = {1085--1090},
  publisher = {ACS Publications},
}

@Article{marston:2008,
  author    = {Johnson, David and Marston, George},
  title     = {The gas-phase ozonolysis of unsaturated volatile organic compounds in the troposphere},
  journal   = {Chem. Soc. Rev.},
  year      = {2008},
  volume    = {37},
  number    = {4},
  pages     = {699--716},
  publisher = {Royal Society of Chemistry},
}

@Article{molpro:2020,
  author  = {Werner, Hans-Joachim and Knowles, Peter J. and Manby, Frederick R. and Black, Joshua A. and Doll, Klaus and Hesselmann, Andreas and Kats, Daniel and Koehn, Andreas and Korona, Tatiana and Kreplin, David A. and Ma, Qianli and Miller, III, Thomas F. and Mitrushchenkov, Alexander and Peterson, Kirk A. and Polyak, Iakov and Rauhut, Guntram and Sibaev, Marat},
  title   = {{The Molpro quantum chemistry package}},
  journal = {J. Chem. Phys.},
  year    = {{2020}},
  volume  = {{152}},
  number  = {{14}},
  pages   = {{144107}},
}

@article{liu:2014,
  title={Infrared-driven unimolecular reaction of CH$_3$CHOO Criegee intermediates to OH radical products},
  author={Liu, Fang and Beames, Joseph M and Petit, Andrew S and McCoy, Anne B and Lester, Marsha I},
  journal={Science},
  volume={345},
  number={6204},
  pages={1596--1598},
  year={2014},
  publisher={American Association for the Advancement of Science}
}

@Article{zhou:2019,
  author    = {Zhou, Xiaohu and Liu, Yiqiang and Dong, Wenrui and Yang, Xueming},
  title     = {Unimolecular reaction rate measurement of \textit{syn}-CH$_3$CHOO},
  journal   = {J. Phys. Chem. Lett.},
  year      = {2019},
  volume    = {10},
  number    = {17},
  pages     = {4817--4821},
  publisher = {ACS Publications},
}

@article{MM.criegee:2021,
  title={Thermal and Vibrationally Activated Decomposition of the
                  \textit{syn}-CH$_3$CHOO Criegee Intermediate},
  author={Upadhyay, Meenu and Meuwly, Markus},
  journal={ACS Earth Space Chem.},
  volume={5},
  number={12},
  pages={3396--3406},
  year={2021},
  publisher={ACS Publications}
}

@Article{welz:2012,
  author    = {Welz, Oliver and Savee, John D and Osborn, David L and Vasu, Subith S and Percival, Carl J and Shallcross, Dudley E and Taatjes, Craig A},
  title     = {Direct kinetic measurements of Criegee intermediate (CH$_2$OO) formed by reaction of CH$_2$I with O$_2$},
  journal   = {Science},
  year      = {2012},
  volume    = {335},
  number    = {6065},
  pages     = {204--207},
  publisher = {American Association for the Advancement of Science},
}

@Article{MM.criegee:2023,
  author    = {Upadhyay, Meenu and Topfer, Kai and Meuwly, Markus},
  title     = {Molecular simulation for atmospheric reactions: non-equilibrium dynamics, roaming, and glycolaldehyde formation following photoinduced decomposition of \textit{syn}-acetaldehyde oxide},
  journal   = {J. Phys. Chem. Lett.},
  year      = {2023},
  volume    = {15},
  number    = {1},
  pages     = {90--96},
  publisher = {ACS Publications},
}

@article{mcgillen:2023,
  title={Ozonolysis can produce long-lived greenhouse gases from commercial refrigerants},
  author={McGillen, Max R and Fried, Zachary TP and Khan, M Anwar H
                  and Kuwata, Keith T and Martin, Connor M and
                  O’doherty, Simon and Pecere, Francesco and
                  Shallcross, Dudley E and Stanley, Kieran M and
                  Zhang, Kexin},
  journal=pnas,
  volume={120},
  number={51},
  pages={e2312714120},
  year={2023},
  publisher={National Academy of Sciences}
}

@article{cox2020evaluated,
  title={Evaluated kinetic and photochemical data for atmospheric
                  chemistry: Volume VII--Criegee intermediates},
  author={Cox, R Anthony and Ammann, Markus and Crowley, John N and
                  Herrmann, Hartmut and Jenkin, Michael E and McNeill,
                  V Faye and Mellouki, Abdelwahid and Troe, J{\"u}rgen
                  and Wallington, Timothy J},
  journal={Atmos. Chem. Phys. Discuss.},
  volume={20},
  pages={13497–13519},
  year={2020},
  publisher={G{\"o}ttingen, Germany}
}

@Article{MM.h2coo:2025,
  author    = {Yin, Cangtao and K{\"a}ser, Silvan and Upadhyay, Meenu and Meuwly, Markus},
  title     = {Photodissociation dynamics of energized H$_2$COO: Formation of molecular products},
  journal   = {J. Chem. Phys.},
  year      = {2025},
  volume    = {163},
  number    = {21},
  pages     = {214305},
  publisher = {AIP Publishing},
}

@article{yin2024revealing,
  title={Revealing new pathways for the reaction of Criegee
                  intermediate CH$_2$OO with SO$_2$},
  author={Yin, Cangtao and Czak{\'o}, G{\'a}bor},
  journal={Comm. Chem.},
  volume={7},
  number={1},
  pages={157},
  year={2024},
  publisher={Nature Publishing Group UK London}
}

@Article{chao2015direct,
  author    = {Chao, Wen and Hsieh, Jun-Ting and Chang, Chun-Hung and Lin, Jim Jr-Min},
  title     = {Direct kinetic measurement of the reaction of the simplest Criegee intermediate with water vapor},
  journal   = {Science},
  year      = {2015},
  volume    = {347},
  number    = {6223},
  pages     = {751--754},
  publisher = {American Association for the Advancement of Science},
}

@Article{yin2018effect,
  author    = {Yin, Cangtao and Takahashi, Kaito},
  title     = {Effect of unsaturated substituents in the reaction of Criegee intermediates with water vapor},
  journal   = {Phys. Chem. Chem. Phys.},
  year      = {2018},
  volume    = {20},
  number    = {30},
  pages     = {20217--20227},
  publisher = {Royal Society of Chemistry},
}

@Article{yin2023nh3,
  author    = {Yin, Cangtao and Czak{\'o}, G{\'a}bor},
  title     = {Full-dimensional automated potential energy surface development and detailed dynamics for the CH$_2$OO+NH$_3$ reaction},
  journal   = {Phys. Chem. Chem. Phys.},
  year      = {2023},
  volume    = {25},
  number    = {39},
  pages     = {26917--26922},
  publisher = {Royal Society of Chemistry},
}

@Article{MM.h2coo:2026,
  author    = {Yin, Cangtao and Meuwly, Markus},
  title     = {Reaction Pathway Dynamics for Atmospheric Decomposition Reactions: Unimolecular Dissociation of H$_2$COO},
  journal   = {J. Phys. Chem. Lett.},
  year      = {2026},
  volume    = {17},
  number    = {24},
  pages     = {6875–6882},
  publisher = {ACS Publications},
}

@Article{xtb.grimme:2021,
  author    = {Bannwarth, Christoph and Caldeweyher, Eike and Ehlert, Sebastian and Hansen, Andreas and Pracht, Philipp and Seibert, Jakob and Spicher, Sebastian and Grimme, Stefan},
  title     = {Extended tight-binding quantum chemistry methods},
  journal   = {WIREs Comput Mol Sci},
  year      = {2021},
  volume    = {11},
  number    = {2},
  pages     = {e1493},
  publisher = {Wiley Online Library},
}

@Article{dmc:1996,
  author  = {Kosztin, Ioan and Faber, Byron and Schulten, Klaus},
  title   = {Introduction to the diffusion Monte Carlo method},
  journal = {Am. J. Phys},
  year    = {1996},
  volume  = {64},
  number  = {5},
  month   = {05},
  pages   = {633--644},
  issn    = {0002-9505},
}

@Article{dmc.bowman:2021,
  author    = {Li, Jeffrey and Qu, Chen and Bowman, Joel M},
  title     = {Diffusion Monte Carlo with fictitious masses finds holes in potential energy surfaces},
  journal   = {Mol. Phys.},
  year      = {2021},
  volume    = {119},
  number    = {17-18},
  pages     = {e1976426},
  publisher = {Taylor \& Francis},
}

@Article{vmd:1996,
  author    = {Humphrey, William and Dalke, Andrew and Schulten, Klaus},
  title     = {VMD: visual molecular dynamics},
  journal   = {J. Mol. Graphics},
  year      = {1996},
  volume    = {14},
  number    = {1},
  pages     = {33--38},
  publisher = {Elsevier},
}

@misc{physnet-tf1,
  howpublished = {\url{https://github.com/MMunibas/PhysNet},},
  note         = {Accessed: 19 August 2026},
}

@misc{physnet-tf2,
  howpublished = {\url{https://github.com/MMunibas/PhysNet-TF2},},
  note         = {Accessed: 19 August 2026},
}

@Article{Lin:2016,
  author  = {Lin, Liang-Chun and Chao, Wen and Chang, Chun-Hung and Takahashi, Kaito and Lin, Jim Jr-Min},
  title   = {Temperature Dependence of the Reaction of \textit{anti}-CH$_3$CHOO with Water Vapor},
  journal = {Phys. Chem. Chem. Phys.},
  year    = {2016},
  volume  = {18},
  number  = {40},
  pages   = {28189--28197},
}

@Article{Lin:2016.2,
  author  = {Lin, Liang-Chun and Chang, Hung-Tzu and Chang, Chien-Hsun and Chao, Wen and Smith, Mica C. and Chang, Chun-Hung and Lin, Jim Jr-Min and Takahashi, Kaito},
  title   = {Competition between {H$_2$O} and {(H$_2$O)$_2$} Reactions with {CH$_2$OO}/{CH$_3$CHOO}},
  journal = {Phys. Chem. Chem. Phys.},
  year    = {2016},
  volume  = {18},
  pages   = {4557--4568},
}

@article{Watson:2025,
  author  = {Watson, Nathan A. I. and Newland, Mike J. and Nelson, Beth S.
             and Rickard, Andrew R. and Beames, Joseph M.},
  title   = {Tropospheric Alkene Ozonolysis Chemistry: An Extended
             Computational Chemistry Assessment of Structural Effects},
  journal = {Environ. Sci.: Adv.},
  year    = {2025},
  volume  = {4},
  pages   = {619--647},
}

@Article{Robinson:2022,
  author  = {Robinson, Callum and Onel, Lavinia and Newman, James and Lade, Rachel and Au, Kendrew and Heard, Dwayne E. and Seakins, Paul W. and Blitz, Mark A. and Stone, Daniel},
  title   = {Unimolecular Kinetics of Stabilized {CH$_3$CHOO} Criegee Intermediates: \textit{syn}-{CH$_3$CHOO} Decomposition and \textit{anti}-{CH$_3$CHOO} Isomerization},
  journal = {J. Phys. Chem. A},
  year    = {2022},
  volume  = {126},
  number  = {39},
  pages   = {6984--6994},
}

@Article{numpy,
  author    = {Harris, Charles R and Millman, K Jarrod and Van Der Walt, St{\'e}fan J and Gommers, Ralf and Virtanen, Pauli and Cournapeau, David and Wieser, Eric and Taylor, Julian and Berg, Sebastian and Smith, Nathaniel J and others},
  title     = {Array programming with NumPy},
  journal   = {Nature},
  year      = {2020},
  volume    = {585},
  number    = {7825},
  pages     = {357--362},
  publisher = {Nature Publishing Group UK London},
}

@InProceedings{networkx,
  author =       {Aric A. Hagberg and Daniel A. Schult and Pieter J. Swart},
  title =        {Exploring Network Structure, Dynamics, and Function using NetworkX},
  booktitle =   {Proceedings of the 7th Python in Science Conference},
  pages =     {11 -- 15},
  address = {Pasadena, CA USA},
  year =      {2008},
  editor =    {Ga\"el Varoquaux and Travis Vaught and Jarrod Millman},
}

@Article{scipy,
  author    = {Virtanen, Pauli and Gommers, Ralf and Oliphant, Travis E and Haberland, Matt and Reddy, Tyler and Cournapeau, David and Burovski, Evgeni and Peterson, Pearu and Weckesser, Warren and Bright, Jonathan and others},
  title     = {SciPy 1.0: fundamental algorithms for scientific computing in Python},
  journal   = {Nat. Methods},
  year      = {2020},
  volume    = {17},
  number    = {3},
  pages     = {261--272},
  publisher = {Nature Publishing Group US New York},
}

@Article{cov.radii:2008,
  author    = {Cordero, Beatriz and G{\'o}mez, Ver{\'o}nica and Platero-Prats, Ana E and Rev{\'e}s, Marc and Echeverr{\'\i}a, Jorge and Cremades, Eduard and Barrag{\'a}n, Flavia and Alvarez, Santiago},
  title     = {Covalent radii revisited},
  journal   = {Dalton Trans.},
  year      = {2008},
  volume    = {37},
  number    = {21},
  pages     = {2832--2838},
  publisher = {The Royal Society of Chemistry},
}

@book{pauling_book:1960,
  title={The nature of the chemical bond and the structure of molecules and crystals: an introduction to modern structural chemistry},
  author={Pauling, Linus},
  volume={18},
  year={1960},
  publisher={Cornell university press},
}

@article{bootstrap:1979,
author = {B. Efron},
title = {{Bootstrap Methods: Another Look at the Jackknife}},
volume = {7},
journal = {Ann. Statist.},
number = {1},
publisher = {Institute of Mathematical Statistics},
pages = {1--26},
year = {1979},
}

@article{marin:2010,
  title={Coincidences in chemical kinetics: Surprising news about
                  simple reactions},
  author={Yablonsky, GS and Constales, Denis and Marin, GB},
  journal={Chem. Eng. Sci.},
  volume={65},
  number={23},
  pages={6065--6076},
  year={2010},
  publisher={Elsevier}
}

@article{vereecken:2012,
  title={Theoretical studies of atmospheric reaction mechanisms in the troposphere},
  author={Vereecken, Luc and Francisco, Joseph S},
  journal=csr,
  volume={41},
  number={19},
  pages={6259--6293},
  year={2012},
  publisher={The Royal Society of Chemistry}
}

@article{cox:2020,
  author  = {Cox, R. Anthony and Ammann, Markus and Crowley, John
                  N. and Herrmann, Hartmut and Jenkin, Michael E. and
                  McNeill, V. Faye and Mellouki, Abdelwahid and Troe,
                  J{\"u}rgen and Wallington, Timothy J.},
  title   = {Evaluated kinetic and photochemical data for atmospheric
                  chemistry: {V}olume {VII} -- {C}riegee
                  intermediates},
  journal = {Atmos. Chem. Phys.},
  volume  = {20},
  number  = {21},
  pages   = {13497--13519},
  year    = {2020},
  doi     = {10.5194/acp-20-13497-2020}
}

@article{MM.nno:2023,
  title={Low-temperature kinetics for the N+ NO reaction: experiment
                  guides the way},
  author={Hickson, Kevin M and San Vicente Veliz, Juan Carlos and
                  Koner, Debasish and Meuwly, Markus},
  journal=pccp,
  volume={25},
  number={20},
  pages={13854--13863},
  year={2023},
  publisher={The Royal Society of Chemistry}
}

\end{document}